\documentclass[11pt]{article}

\usepackage{jheparxiv}

\usepackage{enumitem}

\usepackage[T1]{fontenc} 
\usepackage{hyperref}
\hypersetup{
    colorlinks,
    citecolor=black,
    filecolor=black,
    linkcolor=black,
    urlcolor=black
}
\usepackage[english]{babel}
\usepackage{tikz}
\usepackage{tikz-cd}
\usepackage{tensor}
\usepackage{amsfonts}
\usepackage{amsmath}
\usepackage{amssymb}
\usepackage{amsthm}
\usepackage{mathtools}
\usepackage{tensor}
\usepackage{slashed}
\usepackage{mathrsfs}
\usepackage{bm}
\usepackage{setspace}
\usepackage{amssymb}

\usepackage{bbm}
\usepackage{graphicx}
\usepackage{stmaryrd}
\usepackage{subcaption}
\usepackage{xcolor}
\usepackage{cancel}
\usepackage{bigints}
\newcommand{\scri}{\mathscr{I}}

\renewcommand{\d}{\mathrm{d}}
\newcommand{\C}{\mathbb{C}}
\newcommand{\CP}{\mathbb{CP}}

\newcommand{\M}{\mathbb{M}}
\newcommand{\p}{\partial}
\newcommand{\cO}{\mathcal{O}}

\newcommand{\D}{\mathrm{D}}

\newcommand{\PT}{\mathbb{PT}}

\usepackage{setspace}
\def\bea{\begin{eqnarray}}
\def\eea{\end{eqnarray}}
\newcommand{\beq}{\begin{eqnarray}}
\newcommand{\eqq}{\end{eqnarray}}
 \newcommand{\badat}{\begin{alignedat}}
 \newcommand{\eadat}{\end{alignedat}}

\newcommand{\eal}[1]{\be \begin{aligned} #1 \end{aligned}\end{equation}} 
\newcommand{\eqn}[1]{\be #1 \end{equation}} 
\newcommand{\eqa}[1]{\bea  #1\end{eqnarray}}

\newcommand{\sigmab}{\bar{\sigma}}

\newcommand{\eb}{\bar{\eth}}

\long\def\new#1\endnew{{\bf #1}}		
\long\def\del#1\enddel{}

\def\del{\partial}

\usepackage{xcolor}
\definecolor{oldmauve}{rgb}{0.4, 0.19, 0.28}
\definecolor{pansypurple}{rgb}{0.47, 0.09, 0.29}
\definecolor{burgundy}{rgb}{0.5, 0.0, 0.13}
\definecolor{carminepink}{rgb}{0.92, 0.3, 0.26}
\definecolor{blue(pigment)}{rgb}{0.2, 0.2, 0.6}
\definecolor{darkseagreen}{rgb}{0.56, 0.74, 0.56}
\definecolor{darkspringgreen}{rgb}{0.09, 0.45, 0.27}
\definecolor{ceruleanblue}{rgb}{0.16, 0.32, 0.75}

\newcommand{\be}{\begin{eqnarray}}
\newcommand{\en}{\end{eqnarray}}

\author{}
\numberwithin{equation}{section} 

\begin{document}

\begin{titlepage}
  \thispagestyle{empty}

 \begin{flushright}
 \end{flushright}


  \begin{center}  
{\LARGE\textbf{Celestial $Lw_{1+\infty}$ charges from a twistor action}}


\vskip1cm
Adam Kmec\footnote{\fontsize{8pt}{10pt}\selectfont\ \href{mailto:adam.kmec@maths.ox.ac.uk}{adam.kmec@maths.ox.ac.uk}},
Lionel Mason\footnote{\fontsize{8pt}{10pt}\selectfont\ \href{mailto:Lionel.Mason@maths.ox.ac.uk}{lionel.mason@maths.ox.ac.uk}},
Romain Ruzziconi\footnote{\fontsize{8pt}{10pt}\selectfont\ \href{mailto:Romain.Ruzziconi@maths.ox.ac.uk}{romain.ruzziconi@maths.ox.ac.uk}}, 
Akshay Yelleshpur Srikant\footnote{\fontsize{8pt}{10pt}\selectfont \ \href{mailto:Akshay.YelleshpurSrikant@maths.ox.ac.uk}{akshay.yelleshpur@maths.ox.ac.uk}}
\vskip0.5cm

\normalsize
\medskip

\textit{Mathematical Institute, University of Oxford, \\ Andrew Wiles Building, Radcliffe Observatory Quarter, \\
Woodstock Road, Oxford, OX2 6GG, UK}

\end{center}

\vskip0.5cm

\begin{abstract}

The celestial $Lw_{1+\infty}$ symmetries in asymptotically flat spacetimes have a natural geometric interpretation on twistor space in terms of Poisson diffeomorphisms. Using this framework, we provide a first-principle derivation of the canonical generators associated with these symmetries starting from the Poisson BF twistor action for self-dual gravity. We express these charges as surface integrals over the celestial sphere in terms of spacetime data at null infinity. The connection between twistor space and spacetime expressions at $\mathscr{I}$ is achieved via an integral formula for the asymptotic Bianchi identities due to Bramson and Tod. Finally, we clarify how $Lw_{1+\infty}$ transformations are symmetries of gravity from a phase space perspective by showing the invariance of the asymptotic Bianchi identities. 

\end{abstract}

\end{titlepage}

\setcounter{tocdepth}{1}

\tableofcontents

\section{Introduction}
Finding the symmetries of gravity in asymptotically flat spacetimes has been a long quest with several important turning points. The seminal work of Bondi, van der Burg, Metzner, and Sachs \cite{Bondi:1962px,Sachs:1962zza,Sachs:1962wk} showed that the so-called BMS group, which is an infinite-dimensional enhancement of the Poincaré group with supertranslations, plays a crucial role in the description of radiative spacetimes at null infinity. It was later suggested that the BMS group could itself be enhanced into larger symmetry groups, such as extended \cite{Barnich:2009se,Barnich:2010eb,Barnich:2011mi}, generalized \cite{Campiglia:2014yka,Campiglia:2015yka,Compere:2018ylh}, or Weyl \cite{Freidel:2021fxf} BMS groups (see also \cite{Troessaert:2017jcm, Henneaux:2018cst, Henneaux:2018hdj,Henneaux:2019yax,Fuentealba:2022xsz,Fiorucci:2024ndw} for the corresponding discussions at spacelike infinity). These symmetries have been shown to be of major importance to understand the infrared structure of the gravitational scattering, and the interplay between soft theorems and memory effects \cite{Strominger:2017zoo}. More recently, in the context of celestial holography, it has been shown that a larger group of symmetries with Lie algebra $Lw_{1+\infty}$  emerges from the collinear limit of scattering amplitudes \cite{Guevara:2021abz,Strominger:2021mtt}. The action of these symmetries for flat spacetime is not geometric at null infinity, $\scri$, but can be traced back to Penrose's non-linear graviton construction of self-dual spacetimes from twistor space data \cite{Penrose:1976jq,Penrose:1976js} where they act by local structure preserving diffeomorphisms. The connection between this geometric construction and the collinear limit was presented in \cite{Adamo:2021lrv}. Nevertheless, patterns of the $Lw_{1+\infty}$ symmetries have been identified in the gravitational phase space at null infinity in a series of beautiful works \cite{Freidel:2021dfs,Freidel:2021ytz,Geiller:2024bgf}, including in the expansion of the metric in terms of multipole moments \cite{Compere:2022zdz}. 

An essential ingredient in the identification of the symmetries is the construction of the associated surface charges. Various Noetherian procedures have been developed over the years to tackle this subtle problem and provide a first-principle derivation of these canonical generators \cite{Regge:1974zd,Crnkovic:1986ex,Lee:1990nz,Iyer:1994ys,Barnich:2001jy}. The covariant phase space methods were successfully applied to obtain the BMS charges \cite{Ashtekar:1981bq,Wald:1999wa,Barnich:2011mi,Flanagan:2015pxa,Compere:2018ylh,Henneaux:2018cst,Campiglia:2020qvc,Freidel:2021fxf}. However, the charges of \cite{Freidel:2021ytz,Geiller:2024bgf} associated with the $Lw_{1+\infty}$ symmetries were not derived from first principles, but rather via some recursion relations and symmetries arguments. Truncation of the Bianchi identities was assumed at a certain order to preserve the form of these recursion relations at all orders in the expansion. The obstruction to obtain the celestial $Lw_{1+\infty}$ charges from first principles essentially relies on the non-local action of these symmetries on spacetime data \cite{Mason:2023mti,Donnay:2024qwq}, which makes the direct application of the standard covariant phase space methods obscure. 

In this work, we propose a first-principle derivation of the surface charges associated with the $Lw_{1+\infty}$ symmetries. To do so, we start from the action describing self-dual gravity on twistor space \cite{Mason:2007ct}, where the action of the celestial symmetries becomes local and has a natural geometric interpretation as Poisson diffeomorphisms \cite{Geyer:2014lca,Adamo:2021lrv}. At null infinity, this self-dual phase space agrees with that of conventional gravity. In this framework, the charge algebra for these symmetries is easily derived, by contrast with the spacetime computations which are quite cumbersome and can only be achieved at linear order \cite{Freidel:2021ytz,Geiller:2024bgf,Freidel:2023gue}. Our approach involves several salient features of the covariant phase space methods, such as a treatment of non-integrability and the use of a modified bracket \cite{Barnich:2011mi,Compere:2018ylh,Compere:2020lrt,Fiorucci:2020xto,Adami:2020amw,Adami:2021nnf,Freidel:2021cjp,Freidel:2021dxw}. Using the Bramsom-Tod integral formula \cite{Bramson:1977edc, Tod:2001}, we relate our twistor space expressions for the symplectic structure and the charges in terms of spacetime data. The recursion relations of \cite{Freidel:2021ytz,Geiller:2024bgf}  derive naturally from the integral formula and the on-shell conditions for self-duality implied by the twistor action. In particular, this demonstrates that the truncation of the Bianchi identities postulated in \cite{Freidel:2021ytz,Geiller:2024bgf,Freidel:2023gue} precisely corresponds to the self-dual sector of gravity; essentially they correspond the asymtptotic expansion of Newman's $\mathcal{H}$-space constructed from data at null infinity \cite{Newman:1976gc}.

The self-duality equations are well-known to be an integrable system \cite{Boyer:1985aj, Mason:1991rf,Dunajski:2010zz} and the recursion relations of \cite{Freidel:2021ytz,Geiller:2024bgf,Freidel:2023gue} can be understood to be those arising from the associated recursion operators for the self-duality equations as described in \cite{Dunajski:2000iq} for gravity and \cite{Mason:1991rf} for Yang-Mills.  These recursion relations extend the self-duality equations to an integrable hierarchy, and here its definition is adapted to data at null infinity.  In particular, we see that $Lw_{1+\infty}$ arises by application of these recursion operators to the BMS group.

We also check the consistency of the action of the symmetries on twistor space with the one expected on spacetime and show the invariance of the recursion relations. This clarifies in which sense $Lw_{1+\infty}$ is a symmetry of the gravitational solution space. Finally, we compare our results with those of \cite{Freidel:2021ytz,Geiller:2024bgf} and almost find perfect agreement. The differences occur in the expression of the surface charges for superrotation and higher spin charges. In particular, our derivation offers a new and unambiguous proposal for the expression of the angular momentum at $\mathscr I$ in the presence of radiation.

The paper is organized as follows. In Section \ref{sec:Twistor space action}, we review the self-dual gravity action on twistor space, which is the starting point of our analysis. We also introduce notations and conventions for the rest of the paper. In Section \ref{sec:Asymptotic twistor space and Tod's integral formula}, we discuss the asymptotic twistor space and present the Bramson-Tod integral formula which relates twistor space to data at $\mathscr I$. We show that the spacetime recursion relations automatically follow from this integral transform and the equations of motion on twistor space. In Section \ref{sec:symmetry algebra}, we show that, for a natural gauge fixing adapted to the asymptotic twistor space, the residual gauge transformations on twistor space form the $Lw_{1+\infty}$ algebra. We compute the action of these symmetries on spacetime data and show the invariance of the recursion relations. In Section \ref{sec:Charge algebra}, we apply the covariant phase space methods on the twistor action to derive the charge algebra associated with $Lw_{1+\infty}$. We translate these expressions on spacetime using the Bramson-Tod formula. In Section, \ref{sec:Discussion}, we conclude with some remarks and perspectives. The paper is also completed by several Appendices. Appendix \ref{sec:Recursion relation for $f_s$} provides expressions valid at all spin $s$, Appendix \ref{sec:Symmetry algebra computation} provides details on the algebra computation, Appendix \ref{sec:cech version} presents the Bramson-Tod integral formula in the original Čech picture, instead of the Dolbeault picture adopted in the core of the text, Appendix \ref{sec:BMS Transformations} derives the full BMS transformations from the Bramson-Tod integral formula, and finally, Appendix \ref{sec:ProjForms} provides useful formulae for projective forms used throughout the paper.

\section{Twistor space action}
\label{sec:Twistor space action}

In this section, we introduce the twistor space action describing self-dual gravity \cite{Mason:2007ct}. We also review some basics of twistor space as well as notations and conventions used in the rest of the paper.

\paragraph{Notation and conventions.} Non-projective twistor space is denoted $\mathbb{T}=\C^4$ with coordinates 
\begin{equation}
    Z^A = (\mu^{\dot{\alpha}}, \lambda_{\alpha} ) .
    \label{homogeneous coordinates}
\end{equation} 
Here $\alpha=0,1$ and $\dot\alpha=\dot0,\dot 1$ are standard two-component spinor indices for the anti-self-dual (ASD) and self-dual (SD) spinors respectively.  We use the conventions $\omega^\alpha = \epsilon^{\alpha\beta} \omega_\beta$, $\omega_\alpha = \omega^\beta \epsilon_{\beta \alpha}$ to raise/lower spinor indices, where $\epsilon_{\alpha \beta}$ is the Levi-Civita symbol ($\epsilon_{\alpha \beta} = - \epsilon_{\beta \alpha }$, $\epsilon_{01} = 1 = \epsilon_{\dot{0}\dot{1}}$).   We will use spinor helicity notation for the spinor inner products $$\langle \omega \lambda \rangle = \omega^\alpha \lambda_\alpha, \qquad[\mu \pi] = \mu^{\dot{\alpha}} \pi_{\dot{\alpha}}.$$

In Lorentz signature, the complex conjugation on space-time induces a map from non-projective twistor space $\mathbb{T}$ to its dual $\mathbb{T}^*$ leading to a hermitian inner product given by $$\bar{Z}_{A} = (\bar{\lambda}_{\dot{\alpha}}, \bar{\mu}^\alpha ), \qquad Z^A \bar Z_A=\mu^{\dot\alpha}\bar\lambda_{\dot\alpha}+ \lambda_\alpha\bar\mu^{\alpha}.
$$ 
This is preserved by  $SU(2,2) $, the spin group of the conformal group.

We will also take $Z^A$ to be homogeneous coordinates on the corresponding projective space 
\begin{equation}
   \mathbb{CP}^3 =\{Z^A|Z^A \sim t Z^A,  t\in \mathbb{C}^*\} \, .
\end{equation}
In flat space a point $x$ in Minkowski space $ \mathbb{M}$  is represented in  twistor space as a complex projective line or  Riemann sphere, $X\simeq \CP^1$ via the incidence relation 
\begin{equation}
    \mu^{\dot\alpha}=ix^{\alpha\dot\alpha}\lambda_\alpha \, ,\label{incidence}
\end{equation}
this being two linear equations on the four twistor coordinates, it defines the complex projective line $X$ which has homogeneous coordinates $\lambda_\alpha$.

We take  flat projective twistor space to be the open subset of the complex three-dimensional projective space given by
\begin{equation}
    \mathbb{PT} = \mathbb{CP}^3 - \mathbb{CP}^1\, ,
\end{equation} where $\mathbb{CP}^1 = \{ (\mu^{\dot{\alpha}}, \lambda_{\alpha}) \in \mathbb{CP}^3 | \lambda_{{\alpha}} = 0  \}$ is the line corresponding to the vertex of the light cone at infinity $\scri$.
The quotient from $\C^4$ to $\CP^3$ follows by quotienting along $\Upsilon = Z^A \partial_A$ and $\bar\Upsilon = \bar{Z}^A \bar{\partial}_A$, the holomorphic and anti-holomorphic Euler vector fields, respectively. 
Define a $(p,q)$-form $\omega\in \Omega^{p,q}(m,n)$ on $\mathbb{PT}$ of homogeneity degree $(m,n)$ on the projective space to be a section of $\mathcal{O}(m,n)$ on twistor space by
\begin{equation}
    \Upsilon \lrcorner\,\, \omega = 0 = \bar\Upsilon \lrcorner\,\, \omega  \quad,\quad \mathcal{L}_{\Upsilon}\omega = m\omega \quad , \quad \mathcal{L}_{\bar\Upsilon}\omega = n\omega \, .
\label{homogene}
\end{equation}
In particular, we abuse notation to define the line bundles $\mathcal{O}(n) = \mathcal{O}(n,0)$--here we will be considering mostly smooth rather than holomorphic sections of the line bundle $\mathcal{O}(n)$ in the context of the  Dolbeault picture, by contrast with the Čech picture where this notation would denote the sheaf of holomorphic sections.

 With these notations, the homogeneous measure on the Riemann sphere reads as
\begin{equation}
    D \lambda = \langle \lambda d \lambda \rangle, \qquad  D \bar{\lambda} = [ \bar{\lambda} d \bar{\lambda} ] \, ,
    \label{SphereForms}
\end{equation} 
and these will be used to define homogeneous 1-forms on the lines $X$ corresponding to points of space-time $\M$.
Moreover, the holomorphic volume form on $\PT$ is provided by
\begin{equation}
    D^3 Z = \frac{1}{4!}\epsilon_{ABCD}Z^{A}\d Z^{B}\wedge \d Z^{C}\wedge \d Z^{D} \, ,
    \label{holomorphic top form}
\end{equation} 
where $\epsilon_{ABCD}$ is the Levi-Civita symbol in 4 indices.
The Poincar\'e group  is the subgroup of the conformal group that preserves the infinity twistor
\begin{equation}
    I^{AB} = \begin{pmatrix}
        \epsilon^{\dot\alpha \dot\beta} & 0 \\
        0 & 0
    \end{pmatrix}, \qquad  I_{AB} = \begin{pmatrix}
        0 & 0 \\
        0 & \epsilon^{\alpha \beta}
    \end{pmatrix}\, .
\end{equation}
With this, we can define a Poisson bracket on twistor space through
\begin{equation}
    \{ f, g  \} = I^{AB} \partial_{B}f  \partial_{A} g = \epsilon^{\dot{\alpha}\dot{\beta}} \frac{\partial f}{\partial \mu^{\dot \beta}}  \frac{\partial g}{\partial \mu^{\dot \alpha}}\, ,
\end{equation}
which closes when $f,g$ are 
sections of $\mathcal{O}(2)$. This bracket can be naturally extended to act on forms of various homogeneities via the Lie derivative as follows:  
\begin{equation}
    \{ \omega, \sigma  \} = I^{AB} \mathcal{L}_{\partial_{B}}\omega \wedge \mathcal{L}_{\partial_{A}}\sigma = \epsilon^{\dot{\alpha}\dot{\beta}} \mathcal{L}_{\partial_{\dot \beta}}\omega \wedge \mathcal{L}_{\partial_{\dot\alpha}}\sigma \, .
\label{poisson bracket}
\end{equation}

\paragraph{Deforming the complex structure on twistor space.} The complex structure of flat twistor space $\mathbb{PT}$ corresponds to the splitting of the exterior derivative into its holomorphic and anti-holomorphic parts
\begin{equation}
    \partial = d Z^A \frac{\partial}{\partial Z^A} , \qquad  \bar{\partial} = d \bar{Z}^A \frac{\partial}{\partial{\bar Z}^A}\,.
\end{equation}
A deformation of the complex structure can be expressed as 
\begin{equation}
    \bar \partial \quad \longrightarrow \quad \bar \nabla = \bar \partial + V\ , \qquad V = V^B_A d\bar{Z}^A \otimes \frac{\partial}{\partial Z^B}\, ,
\end{equation}
so that $V \in \Omega^{0,1}(\mathbb{PT} , T^{1,0}\mathbb{PT})$, i.e., an anti-holomorphic $(0,1)$-form on $\PT$, with values on the holomorphic tangent bundle $T^{1,0}\mathbb{PT}$. In general, such a $\bar \nabla$ defines an \emph{almost-complex structure} on twistor space. To be a complex structure it must be integrable, $\bar \nabla^2=0$, which gives the Maurer-Cartan equation
\begin{equation}
    \bar \partial V + \frac{1}{2}[V,V] = 0\, .
    \label{integrability1}
\end{equation}
Here $[\cdot, \cdot]$ is the Schouten–Nijenhuis bracket defined as the commutator between the vector fields combined with the wedge product on the $(0,1)$-forms. As we shall explain in Section \ref{sec:Asymptotic twistor space and Tod's integral formula}, asymptotic twistor space \cite{Hansen:1978jz}, and more generally, the nonlinear graviton \cite{Penrose:1976js}, derives this deformation from a Hamiltonian so as to preserve the Poisson structure \eqref{poisson bracket} arising from the infinity twistor (although the latter does so in a \v Cech framework).  Thus $V=\{h,\cdot\}$ and the deformed complex structure reads as
\begin{equation}
    \bar \nabla = \bar \partial + \{h,\cdot\} \, , 
    \label{covariant derivative twistor}
\end{equation}
where $h \in\Omega^{0,1}(\mathbb{PT},\mathcal{O}(2))$ is the Hamiltonian for the deformation. The integrability condition for $\bar\nabla$ \eqref{integrability1} is then essentially equivalent to
\begin{equation}
    \bar \partial h + \frac{1}{2}\{h,h\} = 0\, .
    \label{integrability condition h}
\end{equation}

\paragraph{Self-dual gravity.} We now introduce the twistor space action for gravity with only self-dual interactions. The Poisson BF theory for self-dual gravity is given by
\begin{equation}
  \boxed{  S[g,h] = \frac{1}{2\pi i}\int_{\mathbb{PT}} g\wedge \left(\bar \partial h + \frac{1}{2}\{h,h\}\right) \wedge D^3 Z\, ,}
  \label{self-dual action}
\end{equation}
where $D^3 Z \in\Omega^{3,0}(\mathbb{PT}, \mathcal{O}(4))$ is the holomorphic top form defined in \eqref{holomorphic top form}, and $g \in \Omega^{0,1}(\mathbb{PT}, \mathcal{O}(-6))$ and $h \in \Omega^{0,1}(\mathbb{PT}, \mathcal{O}(2))$ are the off-shell twistor data for respectively the ASD and SD parts of the dynamical gravitational field. The action is invariant under gauge transformations
\begin{equation}
\begin{split}
    \delta_{(\xi,\phi)} h &= \bar\nabla \xi 
 \, , \\
    \delta_{(\xi,\phi)} g &= \{g,\xi \} + \bar \nabla \phi \, ,
\end{split}
\label{gauge transformations}
\end{equation}
where $(\xi,\phi) \in  \mathcal{O}(2)\oplus  \mathcal{O}(-6)$ are smooth and arbitrary gauge parameters. In these transformations, we are using the bracket \eqref{poisson bracket} as well as the deformed complex structure introduced in \eqref{covariant derivative twistor}. The equations of motion of the action \eqref{self-dual action} are given by 
\begin{equation}
\begin{split}
    \bar \nabla g = \bar \partial g + \{h,g\}  &= 0 \, ,\\
     \bar \partial h + \frac{1}{2}\{h,h\} & = 0 \, .
\end{split}
\label{EOM}
\end{equation}  
One can check explicitly that the equations of motion are invariant under the gauge transformations \eqref{gauge transformations} although it also follows from the corresponding invariance of the Lagrangian. On-shell, it follows from \eqref{EOM} that the field $g$ is $\bar \nabla$-closed and $h$ satisfies the integrability condition \eqref{integrability condition h}. In particular, we note that the linearized version of this second equation is $\bar \nabla \delta h = 0$. Moreover, from \eqref{gauge transformations}, the fields are defined up to $\bar \nabla$-exact terms. In other words, the on-shell field $g$ and linearized perturbation $\delta h$ around $h$ modulo gauge transformations define elements of the cohomology groups for the deformed complex structure denoted by $H^{0,1}_{\bar \nabla}(\PT, \mathcal{O}(-6))$ and $H^{0,1}_{\bar \nabla}(\PT, \mathcal{O}(2))$. 

We can connect these fields to 
conventional descriptions of self-dual gravity via integral formulae in linear theory as follows. The Weyl tensor has a spinorial decomposition into an ASD and SD part
\begin{equation}
    C_{\mu\nu\rho\sigma} = \psi_{\alpha\beta\gamma\delta} \epsilon_{\dot \alpha \dot \beta}\epsilon_{\dot \gamma\dot \delta} + \tilde\psi_{\dot\alpha\dot\beta\dot\gamma\dot\delta} \epsilon_{ \alpha  \beta}\epsilon_{\gamma \delta}\, ,
\label{Weyl tensor}
\end{equation}
where $\psi_{\alpha\beta\gamma\delta}, \tilde\psi_{\dot\alpha\dot\beta\dot\gamma\dot\delta}$ are the ASD and SD fully symmetric Weyl spinors respectively. In linear theory, the Penrose transform relates these to twistor data by 
\begin{equation}
\begin{split}
  g\in    H^{0,1}_{\bar \nabla}(\PT, \mathcal{O}(-6)) &\cong \{\psi_{\alpha\beta\gamma\delta} \,\, \text{on}\,\, \mathbb{M}\vert \nabla^{\alpha\dot \alpha} \psi_{\alpha\beta\gamma\sigma} = 0\} \, ,\\
  \delta h \in H^{0,1}_{\bar \nabla}(\PT, \mathcal{O}(2)) &\cong \{ \tilde\psi_{\dot\alpha\dot\beta\dot\gamma\dot\delta}\,\, \text{on}\,\, \mathbb{M}\vert \nabla^{\alpha\dot\alpha}\tilde\psi_{\dot\alpha\dot\beta\dot\gamma\dot\delta}= 0\}\, .
\end{split}
\end{equation}
The correspondence above is  given explicitly by the integral formulae over the Riemann sphere $X\subset \PT$ given by \eqref{incidence}
\begin{align}
    \psi_{\alpha\beta\gamma\delta} & = \frac{1}{2\pi i}\int_{\mathbb{CP}^{1}} \lambda_{\alpha}\lambda_{\beta}\lambda_{\gamma}\lambda_{\delta} \,\,g\vert_{X}\wedge\D \lambda\, , \qquad g\in H^{0,1}_{\bar \nabla}(\PT, \mathcal{O}(-6))\,,\nonumber \\
    \psi_{\dot\alpha\dot\beta\dot\gamma\dot\delta} & = \frac{1}{2\pi i}\int_{\mathbb{CP}^{1}} \D \lambda\wedge \frac{\p^4\delta h }{\p \mu^{\dot\alpha} \p \mu^{\dot\beta}\p \mu^{\dot\gamma}\p \mu^{\dot\delta}}\vert_{X}\, , \qquad \delta h\in H^{0,1}_{\bar \nabla}(\PT, \mathcal{O}(2))\, .
    \label{Penrosetransfo}
\end{align} 

To work on a curved self-dual background, rather than in linear theory, we must perform the nonlinear graviton construction.  Briefly, the lines $X$ corresponding to points $x\in \M$ become deformed by \eqref{covariant derivative twistor} but nevertheless survive as a 4-complex-dimensional family that forms a generic self-dual space-time $\mathcal{M}$ with points $x\in \mathcal{M}$ corresponding to holomorphic curves $X=\CP^1$ of degree one, but now for the deformed complex structure on curved twistor space, $\mathbb{P}\mathcal{T}$.  

On such a curved background, the first of \eqref{Penrosetransfo} still makes good sense to give linearized anti-self-dual fields.  As such backgrounds admit covariantly constant undotted spinors, there are no Buchdahl conditions, and $\psi_{\alpha\beta\gamma\delta}$ will again satisfy the vacuum Bianchi identities
\begin{equation}
    \nabla^{\alpha\dot\alpha} \psi_{\alpha\beta\gamma\delta}=0\, ,\label{Bianchi}
\end{equation}
where now $\nabla_{\alpha\dot \alpha}$ is the covariant derivative on the SD background.
 For the elements of $H^{0,1}_{\bar \nabla}(\PT, \mathcal{O}(2))$ only a potential modulo gauge description survives from which one can find the SD metric perturbation \cite{Mason:2009afn,Adamo:2022mev} although later it will be sufficient for us to simply use the correspondence with $\scri$.
\section{Asymptotic twistor space and the Bramson-Tod integral formula}
\label{sec:Asymptotic twistor space and Tod's integral formula}

The Penrose transform \eqref{Penrosetransfo} allows us to relate fields on twistor space to fields on space-time. However, those formulae are adapted to representing the fields in the bulk space-time, whereas here in this section, we will use formulae adapted to null infinity originally introduced by Bramson and Tod \cite{Bramson:1977edc, Tod:2001} for $g$ and by Sparling \cite{Eastwood:1982,Sparling:1990} for $h$. In doing so, we introduce the notion of asymptotic twistor space and homogeneous coordinates at $\mathscr I$. Asymptotic twistor space encodes the gravitational data at $\scri$ into a deformed complex structure on twistor space and makes contact between Newman's $\mathcal{H}$-space construction and Penrose's nonlinear graviton construction \cite{Hansen:1978jz}.

\paragraph{Asymptotic twistor space and homogeneous coordinates on $\scri$} To obtain the asymptotic twistor space, we complexify the usual Bondi coordinate $u_B$ at null infinity to obtain $\mathscr I_{\mathbb C} = \mathbb C \times S^2$ (the real $\mathscr I$ can be recovered by just imposing a reality condition on $u$ and making sure that the functions on $\mathscr I_{\mathbb{C}}$ are holomorphic in $u$). Twistor space then fibers over $\PT\rightarrow\scri_\C$.  The use of homogeneous coordinates on $\PT$ then suggests homogeneous coordinates on $\scri$ which have the advantage of manifesting full Lorentz invariance, rather than the usual Bondi coordinates $(u_B,z,\bar z)$ which exhibit $SO(3)$ invariance only \cite{Eastwood:1982}.

Thus, we introduce homogeneous coordinates $\lambda_\alpha$ on the celestial sphere with the usual $z=\lambda_1/\lambda_0$.  The Lorentz group acts in the usual way on the spinor index $\alpha=0,1$, and the reduction from $SL(2,\mathbb C)$ to $SU(2)$ follows from the introduction of a vector $T_{\alpha\dot\alpha}$ satisfying $T^2=2$.  This allows us to define 
\begin{equation}
    \hat \lambda_{\alpha}=T_{\alpha\dot\alpha}\bar \lambda^{\dot\alpha}\, ,
\end{equation}
so that the
round sphere metric is written as
\begin{equation}
    ds^2_{S^2}= \frac{D\lambda D\hat \lambda}{\langle \lambda \hat\lambda\rangle^2}=\frac{dzd\bar z}{(1+|z|^2)^2}\, .
\end{equation}
We lift the Bondi coordinate to a homogeneous coordinate $u$ of weight $(1,1)$ by
\begin{equation}
    u= \langle\lambda\hat\lambda\rangle u_B\, .
\end{equation}
This definition follows from Penrose's strong conformal structure \cite{Penrose:1962ij,Penrose:1986uia} that fixes the ratio $du^2: ds^2_{S^2}$ and a conformal rescaling of the sphere metric to remove the $\langle\lambda\hat\lambda\rangle^2$ factor. It can now be checked from standard transformation laws that $u$ is invariant under Lorentz boosts and that the Lorentz group acts in the normal linear way on $\lambda_\alpha$.

With this, the projection $\PT\rightarrow \scri_\C$ is defined through
\begin{equation}
    p: \PT \to \mathscr{I}_\mathbb{C}, \quad (\mu^{\dot{\alpha}}, \lambda_\alpha) \to (u = [\mu \bar\lambda], \lambda_\alpha , \bar{\lambda}_{\dot{\alpha}})\, ,
\label{projection}
\end{equation} where $u$ is a null complexifed time, and $(\lambda_\alpha , \bar{\lambda}_{\dot{\alpha}})$ the homogeneous coordinates on the celestial sphere. Homogeneous functions on $\mathscr I$ satisfy
\begin{equation}
    f(t\bar t u , t \lambda , \bar t \bar \lambda) = t^m \bar t^n  f( u , \lambda , \bar \lambda) \, ,
\end{equation} and naturally descend from sections of $\mathcal{O}(m,n)$ on twistor space \eqref{homogene} using the projection \eqref{projection}. The homogeneity weights $m,n$ are related to the usual spin and boost weights \cite{Penrose:1984uia} through
\begin{equation}
    s = \frac{m-n}{2}, \qquad w = \frac{m+n}{2}.
\end{equation}
The dual vector fields on the Riemann sphere to the forms \eqref{SphereForms} define the Edth operator and its complex conjugate
\begin{equation}
\eth = \frac{\hat\lambda_{\alpha}}{\langle\hat\lambda\lambda\rangle}\frac{\partial}{\partial \lambda_{\alpha}} \quad , \quad \bar \eth = \frac{\lambda_{\alpha}}{\langle\lambda \hat\lambda\rangle}\frac{\partial }{\partial \hat\lambda_{\alpha}}\, ,
\label{def edth}
\end{equation}
which can act on sections of $\mathcal{O}(m,n)$. The action of the Edth operators changes the weights
\begin{equation}
\begin{split}
    \eth&: \mathcal{O}(m,n) \rightarrow \mathcal{O}(m-2,n) \, ,\\
    \bar\eth&: \mathcal{O}(m,n) \rightarrow \mathcal{O}(m,n-2) \, .
\end{split}
\end{equation} Under the projection map \eqref{projection}, we decompose the twistor coordinate $\mu^{\dot \alpha}$ into the image and kernel of $p$
\begin{equation}
    \mu^{\dot \alpha} = \frac{u\hat{\bar\lambda}^{\dot \alpha}}{\langle\hat\lambda \lambda\rangle} + q\bar\lambda^{\dot \alpha} \quad , \quad q = \frac{[\mu\hat{\bar\lambda}]}{\langle\lambda\hat\lambda\rangle}\, ,
    \label{fiber coordinate}
\end{equation}
where $q\in \mathcal{O}(1,-1)$ is the fiber coordinate over the projection to $\scri$. We make an important distinction between the Edth operator on twistor space and the one on $\scri$
\begin{equation}
\begin{split}
    \partial_{0} &= \left. \frac{\hat \lambda_{\alpha}}{\langle\hat \lambda\lambda\rangle}\frac{\partial}{\partial \lambda_{\alpha}}\right|_{\mu} =  \eth - \bar q \partial_{\bar u}+ \frac{ u}{\langle\hat\lambda\lambda \rangle^{2}}\partial_{q}  \, ,\\
    \bar \partial_{0} &= \left.\frac{\lambda_{\alpha}}{\langle\lambda\hat\lambda\rangle}\frac{\partial}{\partial \hat\lambda_{\alpha}}\right|_{\mu} = \bar \eth - q \partial_{u}+ \frac{\bar u}{\langle\hat\lambda\lambda \rangle^{2}}\partial_{\bar q} \, .
\end{split}
\label{TwistorEdth}
\end{equation} In particular, one can check that $\bar\eth u = 0$, but $\bar\partial_0 u \neq 0$.

In this set-up, the deformation of the complex structure on $\PT$ introduced in \eqref{covariant derivative twistor} can be constructed from the asymptotic shear $\bar{\sigma}(u,\lambda,\bar\lambda)$, which is a section of $\mathcal{O}(1,-3)$, as
\begin{equation}
\boxed{h(\mu,\lambda,\bar\lambda)=\int^u \bar{ \sigma}(u',\lambda,\bar\lambda)du' D \bar \lambda  \, .}
\label{h and sigma}
\end{equation} We will also use the notation $h = h_0 D \bar \lambda$. Here $u=[\mu\bar\lambda]$ and $h$ generates the shear according to
\begin{equation}
    \frac{\p h}{\p \mu^{\dot\alpha}}=\bar\lambda_{\dot\alpha} \bar{\sigma}([\mu\bar\lambda],\lambda,\bar\lambda) D\bar{\lambda} .
\end{equation} One can check that this choice ensures that $h$ satisfies its equation of motion \eqref{EOM}.

\paragraph{Peeling and recursion relations.} See chapter 9 of  \cite{Penrose:1986uia} for full details of the following. In asymptotically flat spacetimes, the large $r$ behaviour of the Weyl tensor in a neighbourhood of $\mathscr I$ is dictated by the peeling theorem. We denote the components of the Weyl tensor \eqref{Weyl tensor} by
\begin{equation}
    \begin{split}
        \Psi_{0} &= C_{\mu\nu\rho\sigma} l^{\mu} m^{\nu} l^{\rho} m^{\sigma} = \psi_{\alpha\beta \gamma \delta} \iota^{\alpha} \iota^{\beta}\iota^{\gamma}\iota^{\delta}, \\
        \Psi_{1} &= C_{\mu\nu\rho\sigma} l^{\mu} n^{\nu} l^{\rho} m^{\sigma}  = \psi_{\alpha\beta \gamma \delta} o^{\alpha} \iota^{\beta}\iota^{\gamma}\iota^{\delta}, \\
        \Psi_{2} &= \frac{1}{2}(C_{\mu\nu\rho\sigma} l^{\mu} n^{\nu} l^{\rho} n^{\sigma} - C_{\mu\nu\rho\sigma} l^{\mu} n^{\nu} m^{\rho} \bar m^{\sigma}) = \psi_{\alpha\beta \gamma \delta} o^{\alpha} o^{\beta}\iota^{\gamma}\iota^{\delta},\\
        \Psi_{3} &= C_{\mu\nu\rho\sigma} l^{\mu} n^{\nu} n^{\rho} \bar m^{\sigma} = \psi_{\alpha\beta \gamma \delta} o^{\alpha} o^{\beta}o^{\gamma}\iota^{\delta},\\
        \Psi_{4} &= C_{\mu\nu\rho\sigma} n^{\mu} \bar m^{\nu} n^{\rho} \bar m^{\sigma} = \psi_{\alpha\beta \gamma \delta} o^{\alpha} o^{\beta}o^{\gamma}o^{\delta}\, ,
    \end{split}
\end{equation} where $(l, n , m, \bar{m})$ is the Newman-Unti co-tetrad \cite{Newman:1962cia} and $(\iota, o)$ is the associated spinor dyad \cite{Newman:1961qr,Geroch:1973am}; thus $\langle\iota o\rangle =1$ and $l^{\alpha\dot\alpha}=o^\alpha\bar o^{\dot\alpha}$ and $l^a\p_a=\p/\p r$, and  $n^{\alpha\dot\alpha}=\iota^\alpha\bar\iota^{\dot\alpha}$ with $n^a\p_a=\p_u$ at $\scri$ and the spinors are parallel propagated along $\p_r$. The peeling theorem   gives  
\begin{equation}
    \Psi_{n} = \frac{\Psi_{n}^0}{r^{5-n}} +\mathcal{O}(r^{n-6})\, ,\qquad  n=0, 1, 2, 3, 4 \, .
\end{equation}
We have the following relation with the asymptotic shear 
\begin{equation}
    \Psi^0_4 = \Ddot{\sigma}, \qquad \Psi^0_3 = \bar\eth \dot{\sigma},
    \label{Bianchi identities}
\end{equation} 
and asymptotic Bianchi identities imply the following time evolution equations 
\begin{equation}
\partial_{u}\Psi_{n}^0 = \bar\eth \Psi_{n+1}^0 +(3-n)\bar\sigma \Psi_{n+2}^0, \qquad n=0,1,2,3 \,. 
\label{Bianchi identities2}
\end{equation}
Using the notations of \cite{Freidel:2021ytz,Geiller:2024bgf}, we denote
\begin{equation}
    \mathcal{Q}_s = \Psi^0_{2-s}, \qquad s= -1, 0, 1, 2\, ,
    \label{identifications}
\end{equation} where $s$ denotes the spin-weight of these Newman-Penrose scalars under Lorentz transformations on the celestial sphere; this suggests a continuation of the series in $s$ which can be motivated as follows.

One can show that the subleading orders in $r$ of $\Psi_1$, $\Psi_2$, $\Psi_3$, $\Psi_4$ are completely fixed by the Bianchi identities once the leading orders $\Psi_1^0$, $\Psi_2^0$, $\Psi_3^0$, $\Psi_4^0$ are given.  By contrast there is no Bianchi identity fixing subleading orders in $\Psi_0$ in $r$ in terms of $\Psi^0_0$ and so its expansion in $r$ involves an infinite tower of independent data. Following \cite{Geiller:2024bgf}, we write the expansion of $\Psi_0$ as
\begin{equation}
    \Psi_{0} = \frac{\Psi^0_{0}}{r^5} + \sum_{s=3}^{\infty}\frac{1}{r^{s+3}} \frac{(-1)^{s}}{(s-2)!}\left(\eth^{s-2}\mathcal{Q}_s+\Phi_{s-2}\right) \, ,
    \label{decomposition subleading}
\end{equation}
where this splitting of the sub-leading terms is chosen so that the evolution equations in $u$ can be cast into the form
\begin{equation}
\partial_{u}\mathcal{Q}_s = \bar\eth \mathcal{Q}_{s-1}+(s+1)\bar\sigma \mathcal{Q}_{s-2}\, ,
\label{recursion1}
\end{equation}
and the infinitesimal extended BMS transformations \cite{Barnich:2013axa,Barnich:2019vzx} are
\begin{equation}
    \delta_{(T,Y,\bar Y)} \mathcal{Q}_s = \left(f\partial_{u}+Y\eth+\bar Y\bar \eth+\frac{3-s}{2}\eth Y + \frac{3+s}{2}\bar\eth \bar Y\right)\mathcal{Q}_s + (s+2)\mathcal{Q}_{s-1}\bar\eth f \, ,
    \label{transfo BMS}
\end{equation}
where $f = T + \frac{u}{2}(\eth Y + \bar \eth \bar Y)$, with $T = T (\lambda, \bar \lambda) \in \mathcal{O}(1,1)$ the supertranslations and $Y (\lambda) \in \mathcal{O}(2,0)$, $\bar Y(\bar \lambda) \in \mathcal{O}(0,2) $ the superrotations. The terms $\Phi_{s-2}$ in \eqref{decomposition subleading} can be shown to be expressible as non-local expressions of the $\mathcal{Q}_s$'s appearing in lower order \cite{Geiller:2024bgf}. Neglecting these terms would yield a truncation of the Bianchi identities for $s \ge 4$ in the full theory of gravity \cite{Freidel:2021ytz}. 

We see that the recursion relations \eqref{recursion1} are valid for any spin $s=-1,0, 1,2, 3 \ldots$ and naturally extend \eqref{Bianchi identities2}, as do the transformation laws \eqref{transfo BMS} under BMS. As we shall explain, both the recursion relations of the $\mathcal{Q}_s$ and the transformation laws will arise naturally from twistor space via a natural extension of the Bramson-Tod integral formula.

\paragraph{Bramson-Tod integral formula and its extension.} The original Bramson-Tod integral formula \cite{Tod:2001}  relates twistor space cohomology classes in $H^{1}(\PT, \mathcal{O}(-6))$ in the \v Cech representation to solutions to the asymptotic Bianchi identities \eqref{Bianchi identities2}.  Here, we reformulate this integral in the Dolbeault framework to connect more directly to the ingredient $g$ in the twistor action \eqref{self-dual action}.  In particular, it extends directly to give the $\mathcal{Q}_s$ providing their origin in Noether's theory. The two pictures are completely equivalent on shell, see for example \cite{griffiths2014principles,kodaira2005}; see Appendix \ref{sec:cech version} for the \v Cech version. 

The integral formula is given by
\begin{equation}
   \boxed{ \mathcal{Q}_s = \frac{1}{2\pi i}\int q^{s+2} \d q \wedge g\vert_{X_{(u,\lambda,\bar\lambda)}}= \frac{1}{2\pi i}\int q^{s+2} g_{1} \d q \d\bar q \, ,}
   \label{Tod formula}
\end{equation}
where $X_{(u,\lambda,\bar\lambda)}$ is the fiber over $\mathscr{I}$ for the projection \eqref{projection}, and where $g\vert_{X_{(u,\lambda,\bar\lambda)}} =g_{1} \d\bar q$, with $g_{1} \in \mathcal{O}(-5,-1)$, denotes the component of the form $g$ along the fiber. Here, the integration takes place on the fiber with coordinates $(q,\bar q)$ as introduced in \eqref{fiber coordinate}. From now on, we will also assume the following holomorphicity condition with respect to $u$:
\begin{equation}
     \hat \lambda^{\alpha} \frac{\partial }{\partial \bar \mu^{\alpha }} \lrcorner\,\, g = 0  \, .
     \label{Gauge1}
\end{equation} As discussed at the beginning of this section, this ensures that, on shell, we can analytically continue our twistor data from $\mathscr I$ to $\mathscr I_{\mathbb C}$. In the next section, we will re-interpret this condition as a gauge-fixing condition on twistor space.

Using the Bramson-Tod formula \eqref{Tod formula}, we now show that the closure condition $\bar \nabla g = 0$, which corresponds to the equation of motion for $g$ given in \eqref{EOM}, is equivalent to the recursion relations 
\begin{equation}
\boxed{\partial_{u}\mathcal{Q}_s = \bar\eth \mathcal{Q}_{s-1}+(s+1)\bar\sigma \mathcal{Q}_{s-2}, \qquad s= -1, 0, 1, 2 \ldots}
\label{recursions}
\end{equation} discussed in the previous paragraph. This justifies the identification of the left-hand side of \eqref{Tod formula} with the spin $s$ momenta discussed in the previous paragraph. Using the Bramson-Tod integral formula, the recursion relations impose a constraint equation on $g_{1}$
\begin{equation}
    \begin{split}
        \partial_{u}\mathcal{Q}_{s}-\bar\eth\mathcal{Q}_{s-1}-(s+1)\bar\sigma \mathcal{Q}_{s-2} &= \frac{1}{2\pi i}\int\left[q^{s+2}\partial_{u} g_{1} - q^{s+1}\bar\eth g_{1}-(s+1)\bar\sigma q^{s} g_{1}\right]\d q\d\bar q \\
        &=\frac{1}{2\pi i} \int q^{s+1}[q\partial_{u}g_{1} -\bar\eth g_{1} + \bar \sigma \partial_{q}g_{1}] \d q\d\bar q \, ,
    \end{split}
\end{equation}
where we have integrated by parts and dropped the boundary terms as the fibers are taken to have no boundary. The LHS is zero if the integrand is a total derivative, hence the condition on $g_1$ is
\begin{equation}
    q \partial_{u} g_{1} - \bar\eth g_{1}+\bar\sigma \partial_{q}g_{1} - \frac{1}{\langle\hat\lambda\lambda \rangle^{2}}\partial_{\bar q}g_{0} = 0\, ,
\end{equation}
where $g_{0}\in \mathcal{O}(-4)$ is an arbitrary function which does no depend on $\bar u$, and the $\langle \hat\lambda \lambda \rangle$ factor was inserted for future convenience. This constraint equation arises naturally from the on-shell condition $\bar \nabla g=0$ for $g$ in an appropriate basis of $(0,1)$-forms on twistor space.  This is  adapted to the projection over $\scri$ by first introducing the variables 
\begin{equation}
    \bar{\mathfrak{q}} = \frac{\langle \bar \mu \hat\lambda \rangle}{\langle \lambda \hat \lambda \rangle^{2}} = \frac{\bar q}{\langle \lambda \hat \lambda \rangle} \quad , \quad \bar u_{B} = \frac{\bar u}{\langle \hat \lambda \lambda \rangle}.
\end{equation}
These have  no anti-holomorphic homogeneity weight (and holomorphic  weights $-2$ and $0$ respectively) and so we have
\begin{equation}
    \begin{split}
        \bar \partial  \bar{\mathfrak{q}} &= \frac{\langle\d\bar\mu\hat\lambda\rangle}{\langle\lambda\hat\lambda\rangle^{2}}-\bar{\mathfrak{q}}\frac{\langle\d\hat\lambda\lambda\rangle}{\langle\hat\lambda\lambda\rangle} + \bar u_{B} \bar e^{0}  \, ,  \\
        \bar \partial  \bar u_{B}&= \frac{\langle \d\bar\mu \lambda\rangle }{\langle \hat \lambda\lambda \rangle} - \bar u_{B}\frac{\langle\d\hat\lambda \lambda\rangle }{\langle\hat\lambda \lambda\rangle}    \, ,\\
        \bar e^{0} &= \frac{\D \bar \lambda}{\langle\lambda \hat\lambda\rangle^{2}}, 
    \end{split}
    \label{Basis}
\end{equation}
where these have  vanishing contraction $\bar\Upsilon\lrcorner \,\, \cdot  = 0$; they are all necessarily $\bar\partial$-closed. We expand our generic (0,1)-form $g$ in this basis
 \begin{equation}
     g = g_{0}\bar e^{0} + g_{\bar{\mathfrak{q}}} \bar \partial  \bar{\mathfrak{q}} + g_{\bar u_{B}} \bar \partial  \bar u_{B}  \, .
 \end{equation}
 Now we impose the axial gauge condition \eqref{Gauge1}
\begin{equation}
    \hat \lambda^{\alpha} \frac{\partial }{\partial \bar \mu^{\alpha }} \lrcorner\,\, g = 0 \quad  \Longrightarrow \quad g_{\bar u_{B}} = 0\, ,
\end{equation}
which will result in holomorphicity in $u$ of the other components when $g$ is on-shell. In this gauge, we find
\begin{equation}
    \bar \partial g = \bar \partial g_{0}\wedge \bar e^{0} + \bar\partial g_{\bar{\mathfrak{q}}}\wedge \bar\partial \bar{\mathfrak{q}}  \, .
\end{equation}
For a general section $F$ of $\mathcal{O}(n)$, $\bar \p F$ can be expanded in the basis \eqref{Basis} as 
\begin{equation}
\begin{split}
    \bar \partial F &= \frac{\partial F}{\partial \bar \lambda_{\dot \alpha}}\d \bar\lambda_{\dot \alpha} + \frac{\partial F}{\partial \bar \mu^{\alpha}}\d \bar \mu^{\alpha} \\
    &= -\bar\partial_{0}F\,\D\bar\lambda +\bar\lambda_{\dot\alpha}\frac{\partial F}{\partial \bar\lambda_{\dot \alpha}}\frac{\langle \d \hat\lambda \lambda \rangle}{\langle\hat\lambda \lambda \rangle} + \partial_{\bar u_{B}}F \frac{\langle \d \bar \mu \lambda\rangle}{ \langle \hat\lambda \lambda\rangle}  + \partial_{\bar{\mathfrak{q}}}F \frac{\langle \d \bar \mu \hat \lambda \rangle}{\langle \lambda \hat\lambda\rangle^{2}}\\
    &= \left(\frac{\bar u}{\langle\hat\lambda \lambda\rangle^{2}}\partial_{\bar q}F-\bar\partial_{0}F\right)\D\bar\lambda+\bar\Upsilon F \frac{\langle \d \hat\lambda \lambda \rangle}{\langle\hat\lambda \lambda \rangle} + \partial_{\bar{\mathfrak{q}}}F \bar\partial \bar{\mathfrak{q}} + \partial_{\bar u_{B}}F \bar\partial \bar u_{B} \\
    &=\left(q\partial_{u}F - \bar\eth F\right)\D\bar\lambda + \partial_{\bar{\mathfrak{q}}}F \bar\partial \bar{\mathfrak{q}} + \partial_{\bar u_{B}}F \bar\partial \bar u_{B} \, ,
\end{split}
\end{equation}
where we have used $\bar \Upsilon F = 0$ and the push-forward of the Edth operator from twistor space to $\scri$ in \eqref{TwistorEdth}.
The non-linear term in the complex structure is given by
\begin{equation}
\begin{split}
    \{h,g\} &= \epsilon^{\dot \beta\dot \alpha}\mathcal{L}_{\partial_{\dot \alpha}}h\wedge \mathcal{L}_{\partial_{\dot \beta}}g  \\
    &=\bar \sigma^{0}\bar\lambda^{\dot \alpha} \frac{\partial g_{\bar{\mathfrak{q}}}}{\partial \mu^{\dot \alpha} }\D \bar\lambda\wedge\bar\partial \bar{\mathfrak{q}}\\
    &= \bar \sigma^{0}\partial_{q}g_{\bar{\mathfrak{q}}}\D \bar\lambda\wedge\bar\partial \bar{\mathfrak{q}}  \, .
\end{split}
\end{equation}
Therefore, the $\bar\nabla$-closure condition in the basis \eqref{Basis} is
\begin{equation}
    \bar \nabla g = \left[q\partial_{u} g_{\bar{\mathfrak{q}}}-\bar \eth g_{\bar{\mathfrak{q}}}+ \bar\sigma^{0}\partial_{q}g_{\bar{\mathfrak{q}}} + \frac{1}{\langle\hat \lambda\lambda\rangle}\partial_{\bar q}g_{0}\right] \D\bar\lambda \wedge \bar\partial \bar{\mathfrak{q}}  + \partial_{\bar u_{B}} g_{\bar{\mathfrak{q}}}\,\bar \partial\bar u_{B} \wedge \bar \partial\bar{\mathfrak{q}} + \partial_{\bar u_{B}} g_{0} \, \bar \partial \bar u_{B}\wedge \bar e^{0} .
\end{equation}
The on-shell condition for $g$ then gives the conditions that the components satisfy
\begin{align}
    &\partial_{\bar u} g_{1} = 0 = \partial_{\bar u}g_{0}\, ,\\
    q \partial_{u} g_{1} - \bar\eth g_{1}&+\bar\sigma \partial_{q}g_{1} - \frac{1}{\langle\hat\lambda\lambda \rangle^{2}}\partial_{\bar q}g_{0} = 0\, ,
\end{align}
where we have made the redefinition $g_{\bar{\mathfrak{q}}} = g_{1}\langle\lambda \hat\lambda \rangle$ to obtain the corrected weighted function in the integral formula. The first conditions give holomorphicity in the $u$-coordinate and are a direct consequence of the axial gauge choice \eqref{Gauge1}. The second gives the condition that if $g$ is $\bar \nabla$-closed, the Bramson-Tod integral formula gives solutions to \eqref{recursions} as desired. 

The integral formula \eqref{Tod formula} can be inverted to give an explicit form of $g$ which is only $\bar\nabla$-closed if the recursion relations \eqref{recursions} are satisfied. The Bramson-Tod representative
is given by
\begin{equation}
    g = -\sum_{n=0}^{\infty}\partial_{u}\mathcal{Q}_{n-2}\frac{\bar q^{n}}{(1+q\bar q)^{n+1}} \D\bar \lambda + \sum_{n=0}^{\infty}\mathcal{Q}_{n-2}\frac{(n+1)\bar q^{n}}{(1+q\bar q)^{n+2}}\d \bar q \, .
\end{equation}

\section{\texorpdfstring{$Lw_{1+\infty}$ symmetry algebra}{Lw(1+infinity) symmetry algebra}}
\label{sec:symmetry algebra}

In \cite{Adamo:2021lrv}, the $Lw_{1+\infty}$ symmetry algebra was geometrically realized as the algebra of Hamiltonians for local holomorphic Poisson diffeomorphisms on the region of twistor space that excludes $\lambda_0=0$ and $\lambda_1=0$.  These are the local symmetries of twistor space in the \v Cech description of Penrose's original nonlinear graviton construction \cite{Penrose:1976js}.  In that discussion, the Hamiltonians for the diffeomorphisms play a role also as \v Cech representatives of $H^1(\mathbb{PT},\cO(2))$ which also define infinitesimal deformations and hence generic self-dual gravitons via the Penrose transform.

Here we take a different approach, deriving the symmetries from those of the self-dual twistor action.  Being a Dolbeault description, these are smooth rather than holomorphic functions, essentially being functions of the six real coordinates of twistor space, rather than the three complex variables as in the previous paragraph.  So we perform various gauge fixings to bring the gauge freedom down to functions of three rather than six variables.  In this way, we will arrive at $Lw_{1+\infty}$. 

Thus, in this section, we perform a residual gauge symmetry analysis on twistor space (see e.g. \cite{Ruzziconi:2019pzd} for a review). We start from the solution space associated with \eqref{EOM} and impose a natural gauge fixing. We then derive the subset of gauge transformations \eqref{gauge transformations} that preserve this gauge fixing and show that they form the $Lw_{1+\infty}$ algebra after using an appropriate bracket. We also compute the transformation of the solution space under the residual gauge transformations and deduce the action of $Lw_{1+\infty}$ symmetries on spacetime using the Bramson-Tod integral formula \eqref{Tod formula}. As in the \v Cech description of \cite{Adamo:2021lrv}, we will see that these gauge transformations correspond to an element  $\delta h \in H_{\bar\nabla}^{0,1}(\mathbb{PT},\cO(2))$ defining a self-dual graviton, now in a Dolbeault representation with the gauge transformation being obtained via an integral formula.

\paragraph{Residual gauge symmetries.} In the first instance, we will choose the following gauge fixing conditions on the forms $(g,h)$ on asymptotic twistor space: 
\begin{align}
    \hat\lambda^{\alpha}\frac{\partial }{\partial \bar\mu^{\alpha}} \lrcorner \,\, g &= 0 \, , \label{gaugecond1}\\ \frac{\partial }{\partial \bar\mu^{\alpha}} \lrcorner \,\, h &= 0 \, .\label{gaugecond2}
\end{align}
These choices are analogous to those in \cite{Woodhouse:1985id} adapted to a fibration of twistor space over Euclidean signature space-time.  Here we use the same general idea, but for the fibration of twistor space over $\scri_\C$.   
As discussed above after \eqref{Gauge1}, the first condition gives the on-shell components of $g$ independence of $\bar u$  that follows from the analytic extension of the fields to complex $u$ in $\scri_\C$.  For homogeneity $\geq -1$ the cohomology up the $\CP^1$ fibers vanishes and so a gauge can be found so that $h$ is trivial up the fibers in the $\p_{\bar q}$ direction.  However,  $H^1(\CP^1, \cO(n))\neq 0$ for $n< -1$ so we cannot ask this of $g$. The condition on  $h$ restricts us to the solution space in which we can hope to interpret the $\D \bar\lambda$ component as coming from the shear as in \eqref{h and sigma} although we will see that it is not quite sufficient. The gauge conditions restrict our gauge transformations in the following way
\begin{align}
        \hat\lambda^{\alpha}\frac{\partial }{\partial \bar\mu^{\alpha}} \lrcorner \,\, \delta_{(\xi,\phi)}g &= 0 \quad \Longrightarrow \quad \partial_{\bar u} \phi = 0  \, ,\\
        \frac{\partial }{\partial \bar\mu^{\alpha}} \lrcorner \,\, \delta_{(\xi,\phi)}h &= 0 \quad \Longrightarrow\quad \frac{\partial \xi}{\partial \bar\mu^{\alpha}} = 0 \, ,\label{homorphicXi}
\end{align} where we used \eqref{gauge transformations}. From this we see that the $\xi$ are independent of $\bar\mu^{\alpha}$ while being arbitrary in $\mu^{\dot \alpha}, \lambda_{\alpha}, \bar\lambda_{\dot \alpha}$ so, taking homogeneities into account, we are now reduced to gauge freedom depending on four variables rather than the three we are aiming for.

In the following, we will therefore also impose the further gauge condition 
\begin{equation}
    \mathcal{L}_{\partial_{q}} h = 0 \, .
    \label{gaugecond3}
\end{equation} Together with \eqref{gaugecond2}, this ensures the compatibility with $\eqref{h and sigma}$ where the shear is seen as a function defined on $\mathscr I_{\mathbb{C}}$ uplifted to twistor space, and therefore does not depend on $q$; the residual gauge freedoms at this point will be the supertranslations and this gauge choice is tantamount to performing the Penrose transform to a potential modulo gauge for the self-dual field. Using \eqref{gauge transformations}, this additional gauge condition \eqref{gaugecond3} implies
\begin{equation}\label{GaugeCond.}
    \partial_{q}(q\partial_{u} -\bar\eth  + \bar\sigma \partial_{q})\xi = 0 \, .
\end{equation} We can solve this formally by expanding $\xi$ as a finite polynomial in $q$
\begin{equation}
    \xi_{s} = \sum_{n=0}^{s+1}\xi_{s,n}q^{n}\, ,
\end{equation}
where $\xi_{s,n} = \xi_{s,n}(u,\lambda,\bar\lambda)$ and the gauge condition \eqref{GaugeCond.} gives the following recursion relations 
\begin{equation} \label{XiRecursion}
    \partial_{u}\xi_{s,n-1} - \bar\eth \xi_{s,n} + (n+1)\bar\sigma \xi_{s,n+1} = 0 \, ,\quad \quad \xi_{s, s+1} := \tau_{s}(\lambda, \bar\lambda) \, .
\end{equation}

These recursion relations can be solved by an integral formula similar to \eqref{Tod formula}
\begin{equation}
    \xi_{s,n} =\frac{1}{2\pi i} \int q^{-n-1}\d q \wedge \delta h_{s}\vert_{X_{(u,\lambda, \bar\lambda)}} \, ,
\end{equation}
where $\delta h_{s} \in H^{0,1}_{\bar\nabla}(\mathbb{PT}, \mathcal{O}(2))$ is an infinitesimal deformation around the background $h$. In this way, the local Hamiltonian vector field generated by $\xi_{s}$ under the Poisson bracket \eqref{poisson bracket} corresponds to the infinitesimal deformation $\delta h_{s}$ around the curved background $h$. The point here is that, like the holomorphic Poisson diffeomorphisms of \cite{Adamo:2021lrv}, $\xi_{s}$ is generically singular on twistor space whereas $\delta h_{s}$, which is gauge equivalent to $\bar \nabla \xi_{s}$, will need to be global and smooth to represent a Dolbeault class.  
However, the integral formula is somewhat implicit and it will be instructive to solve the recursion relations \eqref{XiRecursion} more directly.

To proceed, we consider a polynomial of order $s+1$ in $\mu^{\dot \alpha}$
\begin{equation}
    \tilde \xi_{s} = \frac{1}{(s+1)!}\frac{\partial ^{s+1} \tau_{s}}{\partial \bar\lambda^{\dot \alpha_{1}}\cdots {\partial \bar\lambda^{\dot \alpha_{s+1}}}} \mu^{\dot \alpha_{1}}\cdots \mu^{\dot \alpha_{s+1}}  = \sum_{n=-1}^{s} \frac{u^{s-n}q^{n+1}}{(s-n)!} \bar\eth^{s-n} \tau_{s}\, ,
\label{barXi}
\end{equation}
where $\tau_{s} = \tau_{s}(\lambda_{\alpha}, \bar\lambda_{\dot \alpha})$ and is valued in $\mathcal{O}(1-s,1+s)$ and $s=\{-1,0,1, \dots\}$. When 
$\bar\eth^{s+2}\tau_s=0$ these coincide with the twistor space $Lw_{1+\infty}$ generators given in \cite{Adamo:2021lrv} and this ansatz will be consistent with the wedge condition, but we will work more generally here following \cite{Freidel:2021ytz}. 
This ansatz preserves the two first gauge conditions \eqref{gaugecond1} and \eqref{gaugecond2}, i.e. it satisfies \eqref{homorphicXi}. Furthermore, $\tilde\xi_{s}$ solves \eqref{GaugeCond.} in the absence of shear. Acting with this parameter on the curved background given by $h$, gives 
\begin{equation}
    \delta_{\tilde\xi_s} h = \bar \nabla \tilde\xi_{s} = \left[\bar\sigma \sum_{n=0}^{s}\frac{n+1}{(s-n)!}u^{s-n}q^{n}\bar\eth^{s-n}\tau_{s} - \frac{u^{s+1}}{(s+1)!}\bar\eth^{s+2}\tau_{s}\right]\D\bar\lambda\, ,
\end{equation} where we used $\mathcal{L}_{\p_u} h=\bar\sigma D\bar\lambda$ \eqref{h and sigma} to make the shear appear. As we can see this gives a polynomial in $q$ which is not compatible with the gauge condition \eqref{gaugecond3}. Another way of saying this is that $\tilde\xi_{s}$ does not solve \eqref{GaugeCond.}. The first few examples are 
\begin{equation}
    \begin{split}
        \bar\nabla \tilde\xi_{-1} &= -\bar\eth\tau_{-1}\D\bar\lambda\, ,\\
        \bar\nabla \tilde\xi_{0} &= \left[\tau_{0}\bar\sigma - u\bar\eth^{2}\tau_{0}\right]\D\bar\lambda\, ,\\
        \bar\nabla \tilde\xi_{1} &= \left[(u\bar\eth \tau_{1} + 2q\tau_{1})\bar\sigma - \frac{u^{2}} {2}\bar\eth^{3}\tau_{1}\right]\D\bar\lambda \, ,\\
        \bar\nabla \tilde\xi_{2} &= \left[\left(\frac{u^{2}}{2}\bar\eth^{2} \tau_{2} + 2uq\bar\eth \tau_{2} + 3q^{2}\tau_{2}\right)\bar\sigma - \frac{u^{3}}{6}\bar\eth^{4} \tau_{2}\right]\D\bar\lambda\, ,\\
        \bar\nabla \tilde\xi_{3} &=\left[ \left(\frac{u^3}{6}\bar\eth^{3}\tau_3 + u^{2}q\bar\eth^{2}\tau_{3}+3uq^2\bar\eth\tau_{3}+4q^{3}\tau_{3}\right)\bar \sigma - \frac{u^{4}}{24}\bar\eth^{5}\tau_{3}\right]\D\bar\lambda\, ,
    \end{split}
\end{equation} and the issue arises for $s \ge 1$. To remedy this, we use the linear dependence of $q$ in $ \bar\nabla$ to cancel the highest order in $q$ terms by adding an $\mathcal{O}(2)$ function $f_s$ to $\tilde\xi_{s}$ of a lower order in $q$:
\begin{equation}
    \xi_s = \tilde{\xi}_s + f_s\, .
    \label{decompxi}
\end{equation} We must now do this order by order. For $s=-1, 0, 1,2,3$, this gives the field-depended parameters
\begin{equation}
\begin{split}
f_{-1} &= 0 \,   , \\
f_{0} &= 0 \,  , \\
    f_{1} &=  -2 \tau_1 \partial_{u}^{-1}\bar\sigma \,  ,\\
    f_{2} &=  -3q \tau_2\partial_{u}^{-1}\bar\sigma  - 3\bar\eth(\tau_2\partial_{u}^{-2}\bar\sigma) - 2\bar\eth \tau_2\partial_{u}^{-1}(u\bar\sigma)\, ,\\
    f_{3} &= -4q^2 \tau_{3}\partial_{u}^{-1}\bar\sigma -3q\bar\eth\tau_{3}\partial_{u}^{-1}(u\bar\sigma) - 4q\bar\eth(\tau_{3}\partial_{u}^{-2}\bar\sigma)-\bar\eth^{2}\tau_{3}\partial_{u}^{-1}(u^2\bar\sigma) \\
    & \qquad\qquad \qquad \qquad + 8 \tau_{3}\partial_{u}^{-1}(\bar\sigma\partial_{u}^{-1}\bar\sigma) - 3\bar\eth(\bar\eth\tau_{3}\partial_{u}^{-2}(u\bar\sigma)) - 4\bar\eth^{2}(\tau_{3}\partial_{u}^{-3}\bar\sigma) \, .
\end{split} \label{field dependent f}
\end{equation} We refer to Appendix \ref{sec:Recursion relation for $f_s$} for expressions valid for all spin $s$.

\paragraph{Asymptotic symmetry algebra.}

We now compute the asymptotic symmetry algebra formed by the residual gauge transformations on twistor space. We focus on the $\xi \in \Omega^{0,0}(\mathbb{PT}, \mathcal{O}(2))$, as the parameters $\phi \in \Omega^{0,0}(\mathbb{PT}, \mathcal{O}(-6))$ will be shown to generate trivial gauge transformations in Section \ref{sec:Charge algebra}. The requirement \eqref{gaugecond3} is technically demanding as it yields field dependence in the residual gauge transformations, see \eqref{field dependent f}. However, this feature often arises in the gauge-fixing approach to asymptotic symmetries \cite{Barnich:2010eb,Compere:2019bua,Adami:2020ugu,Ruzziconi:2020wrb,Geiller:2021vpg}, and tools have been developed over the years to deal with these subtleties. In particular, the residual gauge transformations form an algebra, provided one uses a modified Lie bracket \cite{Barnich:2010eb} which takes into account the field dependence of the parameters. In our case, we define the bracket 
\begin{equation}
     \{\xi_s,\xi_{s'}\}_{*} = \{ \xi_s,\xi_{s'}\} + \delta_{\xi_s}\xi_{s'}-\delta_{\xi_{s'}}\xi_s\, ,
     \label{modifiedbracketvectors}
\end{equation} where the first term in the right-hand side is the Poisson bracket on twistor space \eqref{poisson bracket}. The two additional terms take into account the field dependence in $\xi$ and ensure that the bracket closes. Denoting $\xi_s \equiv \xi (\tau_s)$ the expression \eqref{decompxi} with \eqref{barXi}-\eqref{field dependent f}, one can show that 
\begin{equation}
   \boxed{\{\xi (\tau_{s}),\xi(\tau_{s'})\}_{*} = \xi (  (s'+ 1)\tau_{s'} \bar\eth \tau_s  -  (s+1)\tau_s \bar\eth \tau_{s'}) \, .  } 
   \label{Lw1infinity}
\end{equation} The gauge parameter on the right-hand side is again of the form \eqref{decompxi} with \eqref{barXi}-\eqref{field dependent f}. We checked explicitly that the algebra closes for all $s$ at leading order in the fields. Furthermore, we also showed that the algebra closes up to $s=5$ to all orders in the fields. We refer to Appendix \ref{sec:Symmetry algebra computation} for more details on these computations and explicit expressions for all $s$. We conclude that the symmetry algebra of the residual gauge diffeomorphisms $\xi$ is given by the $Lw_{1+\infty}$ algebra. Notice that, in this approach, there is no wedge condition on the parameters, which is in agreement with the phase space discussions on spacetime in \cite{Freidel:2021ytz,Geiller:2024bgf}.\footnote{We note that this algebra only becomes a Lie algebra if we restrict to the wedge as the Jacobi identity would not be satisfied. This is because $Lw_{1+\infty}$ has no Lie subalgebra keeping only $s\geq -1$. On spacetime phase-space one can also realize the generalized BMS algebra. We thank Nicolas Cresto and Laurent Freidel for pointing out this subtlety.}

\paragraph{Action of $Lw_{1+\infty}$.} Let us now compute the action of the above residual gauge transformations on the solution space, and deduce the action of $Lw_{1+\infty}$ on spacetime. Here we focus on the residual gauge transformations associated with $\xi \in \Omega^{0,0}(\mathbb{PT}, \mathcal{O}(2))$ symmetries. The starting point for the transformations of the fields is \eqref{gauge transformations}, together with the explicit form of the gauge parameters \eqref{decompxi} discussed in the previous section. Denoting $\delta_{\xi_s} \equiv \delta_s$, the transformation of $h$ reads as
\begin{equation}
    \begin{split}
        \delta_{-1} h &= -\bar\eth\tau_{-1}\D\bar\lambda \, , \\
        \delta_{0} h &= (\tau_{0}\bar\sigma - u\bar\eth^{2}\tau_{0})\D\bar\lambda \, ,\\
        \delta_{1} h &= \left(2\bar\eth(\tau_{1}\partial_{u}^{-1}\bar\sigma) + u\bar\sigma\bar\eth\tau_{1} - \frac{u^{2}}{2}\bar\eth^{3}\tau_{1}\right) \D\bar\lambda \, , \\
        \delta_{2} h &= \left(\frac{u^2}{2}\bar\sigma\bar\eth^{2}\tau_{2}+ 2\bar\eth(\bar\eth\tau_{2}\partial_{u}^{-1}(u\bar\sigma)) + 3\bar\eth^{2}(\tau_{2}\partial_{u}^{-2}\bar\sigma)-3\tau_{2}\bar\sigma\partial_{u}^{-1}\bar\sigma -\frac{u^{3}}{6}\bar\eth^{4}\tau_{2}\right)\D\bar\lambda \, , \\
        \delta_{3} h &= \left(\frac{u^3}{6}\bar\sigma \bar\eth^{3}\tau_{3}+ \bar\eth(\bar\eth^2\tau_{3}\partial_{u}^{-1}(u^2\bar\sigma)) +3 \bar\eth^{2}(\bar\eth \tau_{3}\partial_{u}^{-2}(u\bar\sigma))- 3\bar\sigma \bar\eth\tau_{3}\partial_{u}^{-1}(u\bar\sigma)\right.\\
        &\quad \quad \left.+4\bar\eth^{3}(\tau_{3}\partial_{u}^{-3}\bar\sigma) -8\bar\eth(\tau_{3}\partial_{u}^{-1}(\bar\sigma\partial_{u}^{-1}\bar\sigma))-4\bar\sigma\bar\eth(\tau_{3}\partial_{u}^{-2}\bar\sigma) - \frac{u^{4}}{24}\bar\eth^{5}\tau_{3}\right)\D\bar\lambda \, ,
    \end{split} \label{transfoh}
\end{equation} and so on. Notice, in particular, that the spin $s=-1$ symmetry is acting non-trivially on $h$. Using \eqref{h and sigma}, we can easily deduce the action on the asymptotic shear as $ \delta_{s} \bar\sigma = \partial_{u}\delta_{s}h_{0}$. We display them explicitly for completeness:
\begin{align}
        \delta_{-1} \bar \sigma &= 0 \, , \nonumber  \\
        \delta_{0} \bar \sigma &= \tau_{0}\partial_{u}\bar\sigma - \bar\eth^{2}\tau_{0} \, , \nonumber \\
        \delta_{1} \bar \sigma &= \left(u\bar\eth\tau_{1}\partial_{u} + 2\tau_{1}\bar\eth + 3\bar\eth\tau_{1}\right)\bar\sigma - u\bar\eth^{3}\tau_{1} \, ,\label{transfosSigma} \\
        \delta_{2} \bar \sigma &= \frac{u^{2}}{2}(\bar\eth^{2}\tau_{2}\partial_{u}\bar\sigma - \bar\eth^{4}\tau_{2})+u(3\bar\eth^{2}\tau_{2}\bar\sigma + 2\bar\eth\tau_{2}\bar\eth\bar\sigma) + 3(\bar\eth^{2}-\partial_{u}\bar\sigma)(\tau_{2}\partial_{u}^{-1}\bar\sigma) - 3\tau_{2}\bar\sigma^{2} \, , \nonumber \\
        \delta_{3} \bar \sigma &= \frac{u^3}{6}(\bar\eth^{3}\tau_{3}\partial_{u}\bar\sigma - \bar\eth^{5}\tau_{3}) + \frac{u^2}{2}(3\bar\eth^{3}\tau_{3} \bar\sigma + 2\bar\eth^{2}\tau_{3}\bar\eth\bar\sigma) + 3u\left((\bar\eth^{2} - \partial_{u}\bar\sigma)(\bar\eth\tau_{3}\partial_{u}^{-1}\bar\sigma )-\tau_{3}\bar\sigma^{2}\right) \nonumber \\
        &  +4\bar\eth^{3}(\tau_{3}\partial_{u}^{-2}\bar\sigma)-3\bar\eth^{2}(\bar\eth\tau_{3}\partial_{u}^{-2}\bar\sigma) - \bar\eth\tau_{3}\partial_{u}\bar\sigma \partial_{u}^{-2}\bar\sigma - 4\tau_{3}\partial_{u}\bar\sigma \partial_{u}^{-2}\bar\eth \bar\sigma - 12 \bar\sigma \bar\eth(\tau_{3}\partial_{u}^{-1}\bar\sigma) - 8\tau_{3}\bar\eth\sigma\partial_{u}^{-1}\bar\sigma \, ,\nonumber
\end{align} and so on. These transformations reproduce correctly those of \cite{Freidel:2021ytz,Geiller:2024bgf}. In particular, the terms linear in the shear agree with the transformations deduced from twistor space in \cite{Donnay:2024qwq}. The spin $s=-1$ symmetry does not act on the shear, which explains why it was invisible in the phase space approach on spacetime \cite{Freidel:2021ytz,Geiller:2024bgf}. The $s=0,1$ transformations reproduce the expected transformations under supertranslations $T$ and superrotations $\bar Y$ by identifying $T= \tau_0$ and $\bar Y = 2\tau_1$ \cite{Barnich:2013axa,Barnich:2019vzx}. 

The transformations of the spin $s$ momenta $\mathcal{Q}_s$ introduced above can be computed similarly, using the Bramson-Tod integral formula \eqref{Tod formula}. Starting from 
\begin{equation}
    \delta_{s'} \mathcal{Q}_s = \frac{1}{2\pi i}\int q^{s+2}\d q \wedge\{g,\xi_{s'}\}\vert_{X_{(u,\lambda,\bar\lambda)}}\, ,
\end{equation} and using again \eqref{gauge transformations}, we find explicitly, on-shell,
\begin{align}
        \delta_{-1} \mathcal{Q}_s &= 0 \, ,\nonumber \\
        \delta_{0} \mathcal{Q}_s &= \tau_{0}\partial_{u}\mathcal{Q}_{s} + (s+2)\mathcal{Q}_{s-1}\bar\eth\tau_{0} \, , \nonumber \\
        \delta_{1} \mathcal{Q}_s &= \left(u\bar\eth \tau_{1}\partial_{u}+ 2\tau_{1}\bar\eth + (s+3)\bar\eth\tau_{1}\right)\mathcal{Q}_{s}+ u(s+2)\mathcal{Q}_{s-1}\bar\eth^{2}\tau_{1} \, , \label{transformationsQs} \\
         \delta_{2} \mathcal{Q}_s &= \frac{u^2}{2}\left(\bar\eth^{2}\tau_{2}\partial_{u}\mathcal{Q}_{s}+(s+2)\mathcal{Q}_{s-1}\bar\eth^{3}\tau_{2}\right)+u\left(2\bar\eth\tau_{s}\bar\eth \mathcal{Q}_{s}+(s+3)\mathcal{Q}_{s}\bar\eth^{2}\tau_{2}\right) \nonumber \\
         & \qquad + 3\tau_{2}\bar\eth\mathcal{Q}_{s+1}+(s+4)\bar\eth\tau_{2} \mathcal{Q}_{s+1} - 3\left(\partial_{u}\mathcal{Q}_{s}+(s+2)\mathcal{Q}_{s-1}\bar\eth\right)\left(\tau_{2}\partial_{u}^{-1}\bar\sigma\right) \, ,\nonumber  \\
        \delta_{3} \mathcal{Q}_s &=\frac{u^{3}}{6}\left(\bar\eth^{3}\tau_{3}\partial_{u}\mathcal{Q}_{s}+(s+2)\mathcal{Q}_{s-1}\bar\eth^{4}\tau_{3}\right)+\frac{u^{2}}{2}\left(2\bar\eth^{2}\tau_{3}\bar\eth\mathcal{Q}_{s}+(s+3)\bar\eth^{3}\tau_{3}\mathcal{Q}_{s}\right)  \nonumber \\
        &\qquad + u\left(3\bar\eth\tau_{3}\bar\eth\mathcal{Q}_{s+1}+(s+4)\bar\eth^{2}\tau_{3}\mathcal{Q}_{s+1}-3\left(\partial_{u}\mathcal{Q}_{s}+(s+2)\mathcal{Q}_{s-1}\bar\eth\right)(\bar\eth\tau_{3}\partial_{u}^{-1}\bar\sigma)\right)  \nonumber \\
        & \qquad + 4\tau_{3}\bar\eth\mathcal{Q}_{s+2} +(s+5)\bar\eth\tau_{3}\mathcal{Q}_{s+2} -4\left(2\bar\eth\mathcal{Q}_{s}+(s+3)\mathcal{Q}_{s}\bar\eth\right)(\tau_{3}\partial_{u}^{-1}\bar\sigma)  \nonumber \\
        &\qquad - \left(\partial_{u}\mathcal{Q}_{s}+(s+2)\mathcal{Q}_{s-1}\bar\eth\right)\left(\bar\eth\tau_{3}\partial_{u}^{-2}\bar\sigma+ 4\tau_{3}\partial_{u}^{-2}\bar\eth\bar\sigma\right) \, .\nonumber
\end{align} We note for both the transformation of $\bar \sigma$ and $\mathcal{Q}_{s}$ there is a recursive structure. In particular, the expressions with a non-zero power of $u$ are built from a lower spin transformation of the same object. The recursive nature is made manifest by expressing a general spin $s$ variations in terms of the $\beta_{n,p}$ found in Appendix \ref{sec:Recursion relation for $f_s$}
\begin{equation}
    \begin{split}
        \delta_{s}\bar\sigma &= \partial_{u}\left(\bar\sigma \sum_{p = 0}^{\lfloor\frac{s}{2}\rfloor}\beta_{1,p} - \sum_{p=0}^{\lfloor\frac{s+1}{s}\rfloor}\bar\eth\beta_{0,p}\right)\, ,\\
        \delta_{s'}\mathcal{Q}_{s} &= \sum_{p=0}^{\lfloor\frac{s'}{2}\rfloor}\left[\beta_{1,p}\partial_{u}\mathcal{Q}_{s} + (s+2)\bar\eth\beta_{1,p}\mathcal{Q}_{s-1}\right] \\
        &\qquad \qquad + \sum_{n=1}^{s'}\sum_{p=0}^{\lfloor\frac{s'-n}{2}\rfloor}\left[(n+1)\beta_{n+1,p}\bar\eth\mathcal{Q}_{s+n-1}+(s+2+n)\bar\eth\beta_{n+1,p}\mathcal{Q}_{s+n-1}\right] \, .
    \end{split}
\end{equation}
Again, the symmetry $s=-1$ does not act on these objects, and the cases $s=0,1$ reproduce the part of the BMS symmetries associated with $T$ and $\bar Y$ (the transformation under $Y$ is deduced in Appendix \ref{sec:BMS Transformations}). To our knowledge, this is the first time the transformations \eqref{transformationsQs} are deduced systematically for all $s$. This illustrates the power of the formula \eqref{Tod formula}. Finally, let us mention that the action of the residual gauge transformations on the solution space forms a representation of the $Lw_{1+\infty}$ symmetries:
\begin{equation}
    [\delta_{\xi},\delta_{\xi'}] = \delta_{\{\xi,\xi'\}_{*}} \, .
\end{equation}
This can be checked explicitly on all the fields using the following relations on twistor space:
\begin{equation}
    [\delta_{\xi},\delta_{\xi'}]g = \{g,\{\xi,\xi'\}_{*}\}\,  ,\qquad [\delta_{\xi},\delta_{\xi'}]h = \bar \nabla \{\xi,\xi'\}_{*}\, ,
\end{equation}
where the Jacobi identity on the twistor space bracket \eqref{poisson bracket} was used. These relations also ensure that the modified bracket \eqref{modifiedbracketvectors} satisfies the Jacobi identity.

\paragraph{Invariance of the recursion relations under $Lw_{1+\infty}$.} We now make the following important observation: the recursion relations \eqref{recursions} are invariant under the action of the $Lw_{1+\infty}$. This is a direct consequence of the fact that the equation of motion for $g$ in \eqref{EOM} is invariant under gauge transformations \eqref{gauge transformations}, combined with the fact that the recursion relations \eqref{recursions} are deduced from the closure condition of $g$ through the Bramson-Tod integral formula \eqref{Tod formula}. As a consistency check, one can also verify explicitly the invariance of \eqref{recursions} using the spacetime transformations found in \eqref{transfosSigma} and \eqref{transformationsQs}. More explicitly, for each $s$, one can show the recursion relations are invariant under the higher spin transformations by lifting to twistor space. We first re-write the recursion relations using the Bramson-Tod integral formula
\begin{equation}
\begin{split}
    \delta_{s'}\left(\partial_{u}\mathcal{Q}_{s}-\bar\eth\mathcal{Q}_{s-1}-(s+1)\bar\sigma\mathcal{Q}_{s-2}\right) &= \frac{1}{2\pi i}\delta_{s'}\int q^{s+1}\left[ q\partial_{u}g_{1}-\bar\eth g_{1} + \bar\sigma \partial_{q}g_{1}\right]\d q \d\bar q\\
    & = \frac{1}{2\pi i\langle \lambda\hat \lambda \rangle}\int q^{s+1}\delta_{s'}\left[\bar\nabla g_{\bar{\mathfrak{q}}}\right]_{\D\bar\lambda}\d q \d\bar q\, ,
\end{split}
\end{equation}
where $\left[\bar\nabla g_{\bar{\mathfrak{q}}} \right]_{\D\bar\lambda}$ denotes the $\D\bar \lambda $ component of the form. The variation of the form is given by
\begin{equation}
\begin{split}
    \delta_{s'} \bar\nabla g_{\bar{\mathfrak{q}}} &= \bar\partial \delta_{s'}g_{\bar{\mathfrak{q}}} + \{\delta_{s'}h , g_{\bar{\mathfrak{q}}}\} + \{h,\delta_{s'}g_{\bar{\mathfrak{q}}}\} \\
    &=\bar\partial\{g_{\bar{\mathfrak{q}}},\xi_{s'}\}+\{\bar\nabla\xi_{s'},g_{\bar{\mathfrak{q}}}\} + \{h,\{g_{\bar{\mathfrak{q}}},\xi_{s'}\}\}\\
    & = \{\bar\partial g_{\bar{\mathfrak{q}}}, \xi_{s'}\} + \{\{h,\xi_{s'}\}, g_{\bar{\mathfrak{q}}}\} + \{\{\xi_{s'},g_{\bar{\mathfrak{q}}}\},h\}\\
    & = \frac{1}{\langle \lambda \hat\lambda\rangle}\{\partial_{\bar q}g_{0}, \xi_{s'}\}\D\bar\lambda + \{\partial_{\bar{\mathfrak{q}}}g_{\bar{\mathfrak{q}}} ,\xi_{s'}\}\bar\partial\bar{\mathfrak{q}}\, ,
\end{split}
\end{equation}
where we have used the equations of motion for $g_{\bar{\mathfrak{q}}}$ and the Jacobi identity for \eqref{poisson bracket}. Finally using the restriction on the gauge parameters \eqref{homorphicXi} due to the gauge conditions, we find
\begin{equation}
\delta_{s'}\left(\partial_{u}\mathcal{Q}_{s}-\bar\eth\mathcal{Q}_{s-1}-(s+1)\bar\sigma\mathcal{Q}_{s-2}\right) = \frac{1}{2\pi i\langle\hat \lambda \lambda \rangle^{2}}\int \partial_{\bar q}\left(q^{s+1}\{g_{0}, \xi_{s'}\}\right)\d q\d \bar q = 0 \, ,
\end{equation}
a total boundary term that vanishes in the integral. Therefore, the higher spin variations preserve the recursion relations
\begin{equation}
   \partial_u \delta_{s'} \mathcal{Q}_s = \bar\eth \delta_{s'} \mathcal{Q}_{s-1} + (s+1) \delta_{s'} \bar \sigma \mathcal{Q}_{s-2} +  (s+1) \bar \sigma \delta_{s'}  \mathcal{Q}_{s-2} \, ,
\end{equation} as desired. This clarifies in which sense $Lw_{1+\infty}$ are symmetries from a spacetime perspective.

\paragraph{Wedge condition}At this stage, as mentioned below \eqref{Lw1infinity}, the wedge condition does not seem to appear in our analysis. However, if one requires the radiative phase space to obey the Schwartzian falloffs as in \cite{Freidel:2021ytz,Geiller:2024bgf,Freidel:2022skz}, 
\begin{equation}
  \lim_{u\to \pm \infty} \bar \sigma  = \mathcal{O} (u^{-s -\epsilon}) \, ,
   \label{Schwartz falloffs}
\end{equation} the $Lw_{1+\infty}$ algebra reduces to its wedge subalgebra, $Lw_{1+\infty}^\wedge$, defined by the commutation relations 
\begin{equation}\label{TauBracket}
     [\tau_s, \tau_{s'}] =\epsilon^{\dot \beta \dot \alpha} \frac{\partial \tau_{s}}{\partial \bar\lambda^{\dot \alpha}} \frac{\partial \tau_{s'}}{\partial \bar\lambda^{\dot \beta}} = (s'+ 1)\tau_{s'} \bar\eth \tau_s  -  (s+1)\tau_s \bar\eth \tau_{s'}\, ,
\end{equation} together with the wedge condition on the parameters $\tau_s (\lambda, \bar \lambda)$
\begin{equation}
    \bar \eth^{s+2} \tau_s = 0 \, .
    \label{wedge condition parameters}
\end{equation} Indeed, preserving \eqref{Schwartz falloffs} under the action of $Lw_{1+\infty}$ implies $\delta_s \bar \sigma|_{\mathscr{I}_{\pm}} = 0$, which requires setting all the non-homogenous terms in the transformation written in \eqref{transfosSigma} to $0$, leading to \eqref{wedge condition parameters}. Notice that the field-dependent contributions in \eqref{decompxi} vanish at $\mathscr{I}^+_\pm$ if we impose the falloffs \eqref{Schwartz falloffs}. Hence, we have $\xi_s|_{\mathscr{I}^+_\pm} = \tilde \xi_s |_{\mathscr{I}^+_\pm}$. Interestingly, using the bracket \eqref{poisson bracket}, the parameters $\tilde \xi_s$ defined in \eqref{barXi} with the wedge condition \eqref{wedge condition parameters} form the $Lw_{1+\infty}^\wedge$ algebra
\begin{equation}\label{WedgeAlgebra}
    \{\tilde \xi (\tau_{s}),\tilde \xi(\tau_{s'})\} = \tilde \xi (  (s'+ 1)\tau_{s'} \bar\eth \tau_s  -  (s+1)\tau_s \bar\eth \tau_{s'})\vert_{\wedge}\, .
\end{equation}
The explicit computations concerning the algebra can be found in Appendix \ref{sec:Symmetry algebra computation}. To connect with the geometric interpretation of $Lw_{1+\infty}^\wedge$ symmetries discussed in \cite{Adamo:2021lrv, Mason:2022hly}, we can expand $\tilde \xi_s$ in the wedge as
\begin{equation}
    \tilde\xi_{s}(\mu^{\dot \alpha},\lambda_{\alpha}) = \sum_{\vert m\vert \leq p-1}\sum_{a \in \mathbb{Z}} \tau_{m,a}^{p} w^{p}_{m,a} \, , \qquad w^{p}_{m,a} : = \frac{(\mu^{\dot 0})^{p+m-1}(\mu^{\dot 1})^{p-m-1}}{2\lambda_{0}^{p-a-2}\lambda_{1}^{p+a-2}}\, ,
\end{equation}
where we define $s = 2p-3$ and $\tau_{m,a}^{p}$ are constants. Using these definitions, the algebra \eqref{WedgeAlgebra} gives the well-known structure constants
\begin{equation}
    \{w^{p}_{m,a}, w^{q}_{n,b} \} = (m(q-1) - n(p-1))w^{p+q-2}_{m+n , a+b}\,.
\end{equation}
As defined, the $\tilde\xi$ can be viewed as Lie algebra homomorphisms, and one can show that the $\tau_{s}$ have a following expansion
\begin{equation}
     \tau_{s}(\lambda_{\alpha}, \bar\lambda_{\dot \alpha}) = \sum_{\vert m\vert \leq p-1}\sum_{a \in \mathbb{Z}} \tau_{m,a}^{p} \tilde w^{p}_{m,a} \, , \qquad \tilde w^{p}_{m,a} : = \frac{(\bar\lambda^{\dot 0})^{p+m-1}(\bar\lambda^{\dot 1})^{p-m-1}}{2\lambda_{0}^{p-a-2}\lambda_{1}^{p+a-2}}\, ,
\end{equation}
where the $\tilde w^{p}_{m,a}$ have the same structure constants under the bracket \eqref{TauBracket}. In the following, we will proceed without imposing the wedge condition restriction \eqref{wedge condition parameters}, and keep the full $Lw_{1+\infty}$ symmetries on the phase space.

\section{Charge algebra}
\label{sec:Charge algebra}

In this section, we construct the symplectic structure on twistor space, from which we deduce the charges associated with the $Lw_{1+\infty}$ symmetries discussed in the previous section. We show that these charges form a representation of these symmetries, provided one uses an appropriate bracket. We then translate these results to spacetime expressions and compare them with previous expressions discussed in the literature.  

\paragraph{Charges on twistor space.} To find the symplectic structure we use the covariant phase space formalism \cite{Crnkovic:1986ex,Lee:1990nz,Iyer:1994ys,Barnich:2001jy}. The variation of the twistor action \eqref{self-dual action} yields
\begin{equation}
\begin{split}
    \delta S =& \frac{1}{2\pi i}\int_{\mathbb{PT}}\D^{3}Z\wedge\left[\left(\bar \partial h + \frac{1}{2}\{h,h\}\right)\wedge \delta g + \bar\nabla g \wedge \delta h\right] \\
    &+ \frac{1}{2\pi i}\int_{\mathbb{PT}}\D^{3}Z\wedge \frac{\partial}{\partial \mu^{\dot \alpha}}\left(g\wedge \epsilon^{\dot \alpha\dot \beta}\frac{\partial h}{\partial \mu^{\dot \beta}}\wedge \delta h\right) +\frac{1}{2\pi i}\int_{\PT}\D^{3}Z \wedge \bar\partial  (  g \wedge \delta h )\, .
\end{split}
\end{equation}
The first term on the right-hand side corresponds to the equations of motion \eqref{EOM}. The second term vanishes due to the gauge fixing condition \eqref{gaugecond2}. The last term is a total derivative term from which the presymplectic potential can be extracted:
\begin{equation}
    \Theta[\delta] = \frac{1}{2\pi i}\int_\Sigma \theta [\delta] =  \frac{1}{2\pi i}\int_{\Sigma}  \delta h \wedge g \wedge  D^3 Z \,.
\end{equation} Here the integration takes place over the co-dimension $1$-surface $\Sigma = \{Z \in \PT |  u_{B} - \bar u_{B} = \text{constant}\}.$\footnote{The space of null twistors is the hypersurface $PN = \{Z \in \PT |  u_{B} - \bar u_{B} = 0\}$.} Taking one more variation of $\Theta$, we obtain the presymplectic form
\begin{equation}
\boxed{\Omega[\delta ,\delta' ]= \frac{1}{2\pi i}\int_{\Sigma} \omega [\delta, \delta'] = \frac{1}{2\pi i}\int_{\Sigma} \delta h \wedge \delta  g \wedge  D^3Z }
\label{symplectictwistor}
\end{equation} with $\omega [\delta, \delta'] = \delta ' \theta [\delta] - \delta \theta [\delta ']$, the presymplectic current.

The charges are then obtained by contracting the presymplectic form with the residual gauge transformations obtained in the previous section. The charge associated with the $\phi$ is given by
\begin{equation}
    \delta_{\phi} \lrcorner \,\,\Omega + \delta F_{\phi} = 0 \, ,  \quad F_{\phi} =  H_{\phi}|^{u_{B} = +\infty}_{u_{B} = - \infty}\,,  \quad H_\phi = \frac{1}{2\pi i}\int_{\partial \Sigma} D^3Z  \wedge h \phi\, .
\end{equation}
However, since $\partial \Sigma$ is the hypersurface in twistor space keeping $u_{B}$ constant and real, in the gauge we are working, $H_{\phi} = 0$. This is because when pulled back to $\partial \Sigma$ the holomorphic top form $D^{3}Z$ expressed in the fibration coordinates contains a $\D\bar\lambda$, see Appendix \ref{sec:ProjForms} for more details. In this case, the symplectic form is degenerate on the phase space because $\delta_{\phi}$ is a non-zero vector in general. Moreover, 
\begin{equation}
    \delta_{\xi} \lrcorner \,\,\Omega + \delta F_{\xi} = 0\, ,  \quad F_{\xi} =  H_{\xi}|^{u_{B} = +\infty}_{u_{B} = - \infty}\,,  \quad  H_\xi = \frac{1}{2\pi i}\int_{\partial \Sigma} D^3Z  \wedge \xi g \, ,
\end{equation} where we have dropped the boundary terms associated with total $\partial_{q}$ terms because we assume the fibers have no boundary throughout. There is also a total $\partial_{u}$ term, which read explicitly as
\begin{multline}
    \frac{1}{2\pi i}\int_{\Sigma} \D^{3}Z \wedge \partial_{u}\left(\partial_{q}\xi_{s} g\wedge \delta h \right) = \\
    \left[\sum_{n=0}^{s}(n+1)\int_{\mathbb{CP}^{1}} \delta h_{0} \mathcal{Q}_{n-2}\left(\frac{u^{s-n}}{(s-n)!}\bar\eth^{s-n}\tau_{s} +\alpha_{n+1}[\bar\sigma,\tau_{s}] \right) \D\lambda\D\bar \lambda\right]_{u_{B}=-\infty}^{u_{B} = +\infty}\, .
\label{chargesXi}
\end{multline} 
This corner term vanishes for $s=0,1$ if we take the same falloffs in $u$ as in \cite{Donnay:2021wrk,Donnay:2022hkf}. Furthermore, this term vanishes for all $s$ if we restrict ourselves to the Schwartzian falloffs for the shear \eqref{Schwartz falloffs}, as considered in \cite{Freidel:2021ytz,Geiller:2024bgf}. We proceed by neglecting this boundary term, but without necessarily assuming the wedge condition.

We see from \eqref{chargesXi} that the charges associated with $\xi \in \Omega^{0,0}(\PT, \mathcal{O}(2))$ are non-vanishing, hence confirming that the $Lw_{1+\infty}$ are indeed large gauge transformations on twistor space. In particular, the charge associated with $s=-1$ does not vanish, which means that the corresponding symmetry has a non-trivial action on phase space. Furthermore, one can easily show that the charges form a representation of the $Lw_{1+\infty}$ symmetries, i.e.
\begin{equation}
   \boxed{ \delta_\xi F_{\xi'} = F_{\{ \xi, \xi'\}_*}} 
\label{algebra fluxes twistor}
\end{equation} where the bracket for the residual gauge parameters on the right-hand side was defined in \eqref{modifiedbracketvectors}. This important result could only be obtained at linear order from a spacetime approach \cite{Freidel:2021ytz,Geiller:2024bgf}, demonstrating the power of the twistor methods presented here. Notice that restricting \eqref{algebra fluxes twistor} to $s=0,1$ reproduces the BMS flux algebra, in agreement with \cite{Donnay:2021wrk,Donnay:2022hkf}. 

It is also instructive to derive the surface charge expressions at finite value of $u_B$, which will coincide with the surface charges integrated on a finite cut of $\mathscr I$ after translation to spacetime, see next paragraph. We have \cite{Lee:1990nz,Iyer:1994ys,Barnich:2001jy}
\begin{equation}
    \delta_\xi  \lrcorner \,\,\omega = d k_\xi, \qquad \cancel{\delta}H_{\xi} = \frac{1}{2\pi i}\int_{\partial \Sigma} k_\xi
\end{equation} where $k_\xi$ is a co-dimension $2$ form on twistor space, which can be integrated over $\partial \Sigma = \{Z \in \PT |  u_{B} - \bar u_{B} = 0, u_{B}=\text{constant}\}$. The resulting charges are non-integrable (i.e.\ non $\delta$-exact on the solution space) and can be split as
\begin{equation}
    \cancel{\delta}H_{\xi} = \delta H_{\xi} - \Xi_{\xi}[\delta]   \,,
    \label{split integrable}
\end{equation}
where the integrable and non-integrable pieces are given by
\begin{equation}
  \boxed{H_{\xi} = \frac{1}{2\pi i}\int_{\partial \Sigma}  D^3 Z\wedge \xi g}   \, , \qquad  \Xi_{\xi}[\delta] = H_{\delta \xi} + \frac{1}{2\pi i}\int_{\partial\Sigma} \underline{\xi}\lrcorner\, \theta[\delta]\,.
   \label{explicit expression}
\end{equation}
where $\underline{\xi} = \{\xi, \,\cdot\,\}$ is the Hamiltonian vector field generated by $\xi$. The explicit expression for the second term in $\Xi_\xi [\delta]$ can be read of from \eqref{chargesXi}. The integrable/non-integrable split adopted here corresponds to the Noetherian split discussed e.g. in \cite{Freidel:2021cjp}. The non-integrable piece $\Xi_\xi$ vanishes when $u \to \pm \infty$ for $s=0,1$ if we take the same falloffs in $u$ as in \cite{Donnay:2021wrk,Donnay:2022hkf}, and vanishes for all $s$ if we restrict ourselves to the Schwartzian falloffs for the shear \eqref{Schwartz falloffs}, as considered in \cite{Freidel:2021ytz,Geiller:2024bgf}. At finite $u$, the non-integrability of the charges appears for $s\ge 1$. This might have been expected, as it is well known that the charges associated to the BMS symmetries in asymptotically flat spacetimes are already non-integrable \cite{Ashtekar:1981bq,Wald:1999wa,Barnich:2011mi,Flanagan:2015pxa,Compere:2018ylh,Henneaux:2018cst,Campiglia:2020qvc,Freidel:2021fxf}. Several tools have been developed to deal with this non-integrability, including a modified bracket \cite{Barnich:2011mi,Compere:2018ylh,Compere:2020lrt,Fiorucci:2020xto,Adami:2020amw,Adami:2021nnf,Freidel:2021cjp,Freidel:2021dxw} defined as
\begin{equation}
    \{H_{\xi}, H_{\xi'}\}_{\star} = \delta_{\xi}H_{\xi'} - \Xi_{\xi}[\delta_{\xi'}]\, .
\end{equation} Using this bracket, one can easily show that 
\begin{equation}
   \boxed{\{H_{\xi}, H_{\xi'}\}_{\star} = H_{\{\xi,\xi'\}_{*}}  \, , }
   \label{charge algebra}
\end{equation} where the bracket for vector fields was given in \eqref{modifiedbracketvectors}. Hence, the charges form a representation of the $Lw_{1+\infty}$ symmetries on twistor space. The consistency of this bracket with Jacobi identities follows directly from the consistency of the bracket \eqref{modifiedbracketvectors}. Using the above arguments, one can check that the result \eqref{charge algebra} consistently reduces to \eqref{algebra fluxes twistor} when taking $u \to \pm \infty$ and assuming the appropriate falloff conditions in $u$. Finally, notice that a different split between integrable and non-integrable pieces in \eqref{split integrable}-\eqref{explicit expression}, 
\begin{equation}
    H_\xi \to H'_\xi = H_\xi + N_\xi, \qquad \Xi_\xi \to \Xi'_\xi = \Xi_\xi + \delta N_\xi\, ,
\end{equation} would not affect $ \cancel{\delta}H_{\xi}$, but it would produce a (trivial) field-dependent 2-cocycle in the right-hand side of \eqref{charge algebra}:
\begin{equation}
    \{H'_{\xi}, H'_{\xi'}\}_{\star} = H'_{\{\xi,\xi'\}_{*}} + K'_{\xi,\xi'}, \qquad K'_{\xi,\xi'} = \delta_\xi N_{\xi'} - \delta_{\xi'} N_{\xi} - N_{\{\xi,\xi'\}_{*}}\, .
\end{equation} This justifies a posteriori why the natural split chosen in \eqref{split integrable}-\eqref{explicit expression} is the most appropriate.

\paragraph{Charges on spacetime.} We now translate the symplectic structure and charges obtained from first principles on twistor space to spacetime expressions using the dictionary discussed in section \ref{sec:Asymptotic twistor space and Tod's integral formula}. Starting from the symplectic structure \eqref{symplectictwistor} and using the Bramson-Tod integral formula \eqref{Tod formula}, we find 
\begin{equation}
\begin{split}
    \Omega & = \frac{1}{2\pi i}\int_{\Sigma} \delta h_{0} \delta g\wedge \d q \wedge \d u \wedge \D\lambda \wedge\D\bar\lambda = -\int_{\mathscr{I}} \delta h_{0} \delta \Psi_{4}^0  \d u \D\lambda \D\bar\lambda\, .
\end{split}
\end{equation} Hence, on the radiative phase space coming from twistor space,  the natural symplectic pairing is between $h_0$ and $\Psi^0_4$. The standard Ashtekar-Streubel symplectic structure on the radiative phase space \cite{Ashtekar:1981bq} can be obtained by integrating by parts with respect to $u$, using \eqref{Bianchi identities} and \eqref{h and sigma}, and assuming strong fall-offs on the news to get rid of the boundary term:
\begin{equation}
    \Omega = \int_{\mathscr{I}} \delta \bar\sigma \delta \dot \sigma   \d u \D\lambda \D\bar\lambda\, .
\end{equation} 
In this case, the symplectic pairing is between the shear and the news. The virtue of choosing the symplectic pairing $(h_0,\Psi^0_4)$ instead of $(\bar\sigma,\dot \sigma)$ is that it allows us to see the non-trivial action of the $s=-1$ symmetry on phase space (compare \eqref{transfoh} with \eqref{transfosSigma}). 

Let us now focus on the integrable part of the charges \eqref{explicit expression} associated with the $Lw_{1+\infty}$ symmetries. Again, using \eqref{Tod formula}, as well as the explicit expressions for the spin $s$ parameters \eqref{decompxi} together with \eqref{barXi}-\eqref{field dependent f}, we find explicit expressions for the surface charges at a cut $\mathcal{S} = \{(u,z,\bar z) \in \mathscr I | u=\text{constant} \}$ of null infinity: 
\begin{equation}
\boxed{\begin{split}
    H_{-1} &= \int_{\mathcal{S}} \tau_{-1} \mathcal{Q}_{-1}\D\lambda\D\bar\lambda \, , \\
    H_{0} &= \int_{\mathcal{S}} \tau_{0} (\mathcal{Q}_{0} - u\bar\eth\mathcal{Q}_{-1})\D\lambda\D\bar\lambda  \, , \\
    H_{1} & = \int_{\mathcal{S}} \tau_{1}\Big(\mathcal{Q}_{1}-u\bar\eth \mathcal{Q}_{0}+\frac{u^2}{2}\bar\eth^{2}\mathcal{Q}_{-1}-2\mathcal{Q}_{-1}\partial_{u}^{-1}\bar\sigma \Big)\D\lambda\D\bar\lambda   \, ,\\
     H_{2} & = \int_{\mathcal{S}} \tau_{2}\Big(\mathcal{Q}_{2} - u\bar\eth\mathcal{Q}_{1}+\frac{u^2}{2}\bar\eth\mathcal{Q}_{0}-\frac{u^3}{6}\bar\eth^3\mathcal{Q}_{-1} \\
     &\qquad\qquad\quad - 3\mathcal{Q}_{0}\partial_{u}^{-1}\bar\sigma+3\bar\eth\mathcal{Q}_{-1}\partial_{u}^{-2}\bar\sigma+2\bar\eth\left(\mathcal{Q}_{-1}\partial_{u}^{-1}(u\bar\sigma)\right)\Big)\D\lambda\D\bar\lambda  \, , \\
     H_3 &= \int_{\mathcal{S}}\tau_{3}\Big(\mathcal{Q}_{3}-u\bar\eth\mathcal{Q}_{2}+\frac{u^{2}}{2}\bar\eth^{2}\mathcal{Q}_{1}-\frac{u^{3}}{6}\bar\eth^{3}\mathcal{Q}_{0}+\frac{u^{4}}{24}\bar\eth^{4}\mathcal{Q}_{-1}-4\mathcal{Q}_{1}\partial_{u}^{-1}\bar\sigma\\
     &\qquad\qquad\quad + 3\bar\eth\left(\mathcal{Q}_{0}\partial_{u}^{-1}(u\bar\sigma)\right) +4\bar\eth\mathcal{Q}_{0}\partial_{u}^{-2}\bar\sigma +8\mathcal{Q}_{-1}\partial_{u}^{-1}(\bar\sigma\partial_{u}^{-1}\bar\sigma) \\
     &\qquad\qquad\quad -3\bar\eth\left(\bar\eth \mathcal{Q}_{-1}\partial_{u}^{-2}(u\bar\sigma)\right)-\bar\eth^{2}\left(\mathcal{Q}_{-1}\partial_{u}^{-1}(u^2\bar\sigma)\right)-4\bar\eth\mathcal{Q}_{-1}\partial_{u}^{-3}\bar\sigma\Big)\D\lambda \D\bar\lambda\, \, .
\end{split}}\label{surface charges spacetime}
\end{equation} 
Using the recursion relations presented in Appendix \ref{sec:Recursion relation for $f_s$}, a general expression for the charges valid for any spin $s$ can be presented:
\begin{equation}
\begin{split}
    H_{s} &= \sum_{n=-1}^{s}\int_{\mathcal{S}} \left(\tau_{s}\frac{(-u)^{s-n}}{(s-n)!}\bar\eth^{s-n}\mathcal{Q}_{n} +\sum_{p=1}^{\lfloor \frac{s-n}{2} \rfloor}\beta_{n+1,p}[\tau_{s},\bar\sigma]\mathcal{Q}_{n}\right)\D\lambda\D\bar\lambda\\
    & = \sum_{n=-1}^{s}\sum_{p=0}^{\lfloor \frac{s-n}{2} \rfloor} \int_{\mathcal{S}}\beta_{n+1,p}[\tau_{s},\bar\sigma]\mathcal{Q}_{n}\D\lambda\D\bar\lambda\, ,
\end{split}
\end{equation}
where the $\beta_{n,p}$ are depended on the coordinates of $\mathcal{S}$, the transformation parameter $\tau_{s}$, and the shear $\bar\sigma$. The $\beta_{n,p}$ are defined recursively in \eqref{eq:precursion}.

Let us make a couple of comments concerning these surface charge expressions. First, we observe that these charges are very similar to those constructed heuristically in \cite{Freidel:2021ytz,Geiller:2024bgf}. By contrast with these works, we find a non-vanishing spin $s=-1$ charge on spacetime, involving the momentum $\mathcal{Q}_{-1} = \Psi^0_3$. This is consistent with the fact that the spin $s=-1$ acts non-trivially on the phase space variable $h$. Notice that this charge would disappear if the wedge condition \eqref{wedge condition parameters} was imposed, see \eqref{Bianchi identities} (in that case, the symmetry $s=-1$ would be a central term). We find a similar contribution $\frac{(-u)^{s+1}}{(s+1)!}\mathcal{Q}_{-1}$, for all spin $s$ charge, which differs from \cite{Freidel:2021ytz,Geiller:2024bgf} and which would disappear on the wedge. Moreover, we find some terms $\sim \partial_u^{-1} \bar \sigma$ of difference compared to \cite{Freidel:2021ytz,Geiller:2024bgf}. Despite these differences, one can verify that our charges \eqref{surface charges spacetime} satisfy the two following criteria: 

\noindent $(i)$ The charges are conserved in time in the absence of radiation, i.e. 
\begin{equation}\label{NoRadiation}
    \Psi^0_4 = 0 \quad \Longrightarrow \quad \frac{d}{du}H_s = 0 \, .
\end{equation} Notice that the non-radiative condition we are considering here is weaker than that in \cite{Freidel:2021ytz,Geiller:2024bgf} where the additional condition $\Psi^0_3 = 0$ is imposed, but it matches exactly the one proposed earlier in \cite{Freidel:2021qpz}. This makes a lot of sense regarding our construction: the canonical variable $\Psi^0_4$ controls the radiation which spoils the conservation of the $Lw_{1+\infty}$ charges constructed out of $\mathcal{Q}_s$, with $s= -1,0,1,2, \ldots$ The explicit expressions of the local fluxes $\mathcal{F}_s = \partial_u H_s$ read as
\begin{align}
        \mathcal{F}_{-1} &= \int_{\mathcal{S}} \tau_{-1}\bar\eth\Psi_{4}^{0}\D\lambda\D\bar\lambda \, , \nonumber \\
        \mathcal{F}_{0} &=\int_{\mathcal{S}}\tau_{0}\left(\bar\sigma\Psi_{4}^{0} - u\bar\eth^{2}\Psi_{4}^{0}\right)\D\lambda\D\bar\lambda \, ,  \nonumber \\
        \mathcal{F}_{1} &=\int_{\mathcal{S}}\tau_{1}\left(\frac{u^{2}}{2}\bar\eth^{3}\Psi_{4}^{0}-u\bar\eth(\bar\sigma\Psi_{4}^{0})-2\bar\eth\Psi_{4}^{0}\partial_{u}^{-1}\bar\sigma\right)\D\lambda\D\bar\lambda \, ,  \\
        \mathcal{F}_{2} &=\int_{\mathcal{S}}\tau_{2}\left(\frac{u^2}{2}\bar\eth^{2}(\bar \sigma \Psi_{4}^{0}) - \frac{u^{3}}{6}\bar\eth^{4}\Psi_{4}^{0} - 3\Psi_{4}^{0}\bar\sigma\partial_{u}^{-1}\bar\sigma + 2\bar\eth\left(\bar\eth \Psi_{4}^{0} \partial_{u}^{-1}(u\bar\sigma)\right)\right)\D\lambda\D\bar\lambda \, ,  \nonumber\\
        \mathcal{F}_{3} &= \int_{\mathcal{S}}\tau_{3}\left(\frac{u^{4}}{24}\bar\eth^{5}\Psi_{4}^{0} - \frac{u^{3}}{6}\bar\eth^{3}(\bar\sigma\Psi_{4}^{0})-\bar\eth^{2}\left(\bar\eth\Psi_{4}^{0}\partial_{u}^{-1}(u^2\bar\sigma)\right)-3\bar\eth\left(\bar\eth^{2}\Psi_{4}^{0}\partial_{u}^{-2}(u\bar\sigma)\right)\right.   \nonumber\\
        &\qquad \qquad  +\left. 3\bar\eth\left(\Psi_{4}^{0}\bar\sigma\partial_{u}^{-1}(u\bar\sigma)\right)+4\bar\eth(\bar\sigma\Psi_{4}^{0})\partial_{u}^{-2}\bar\sigma+8\bar\eth\Psi_{4}^{0}\partial_{u}^{-1}(\bar\sigma\partial_{u}^{-1}\bar\sigma)-4\bar\eth^{3}\Psi_{4}^{0}\partial_{u}^{-3}\bar\sigma \right)\D\lambda\D\bar\lambda\, . \nonumber
  \end{align} 
Using the recursion relations found in Appendix \ref{sec:Recursion relation for $f_s$}, one can further show that the fluxes have a compact expression in terms of the $\beta_{n,p}$
\begin{equation}
    \mathcal{F}_{s} = \int_{\mathcal{S}}\Psi_{4}^{0}\left(\bar\sigma \sum_{p=0}^{\lfloor \frac{s}{2} \rfloor}\beta_{1,p}[\tau_{s},\bar\sigma] - \sum_{p=0}^{\lfloor \frac{s+1}{2} \rfloor} \bar\eth \beta_{0,p}[\tau_{s},\bar\sigma]\right)\D\lambda\D\bar\lambda\, ,
\end{equation}
showing that the charges $H_{s}$ satisfy \eqref{NoRadiation}.

\noindent $(ii)$ One can easily show that the integrated fluxes $\int_{-\infty}^{+\infty} du \mathcal{F}_s$ are exactly the same as those considered in \cite{Freidel:2021ytz,Geiller:2024bgf}, provided one assumes the same falloffs in $u$ on the shear given in \eqref{Schwartz falloffs} allowing for integration by parts. In particular, these fluxes are consistent with the action of the $Lw_{1+\infty}$ symmetries on the radiative phase space.

Let us emphasize that the phase space analysis on twistor space also provides new proposals for the more standard BMS charges. Introducing the supertranslation $T = \tau_0$ and superrotation $\bar Y = 2 \tau_1$ parameters, and using the correspondence \eqref{identifications}, we find the BMS surface charges:\footnote{In our conventions, $4\pi G = 1$.}
\begin{equation}
    \begin{split}
        H_{0} &= \int_{\mathcal{S}} T (\Psi^{0}_{2} - u\bar\eth\Psi^{0}_{3})\D\lambda\D\bar\lambda   \, ,\\
    H_{1} & = \frac{1}{2}\int_{\mathcal{S}} \bar Y\left(\Psi^{0}_{1}-u\bar\eth \Psi^{0}_{2}+\frac{u^2}{2}\bar\eth^{2}\Psi^{0}_{3}-2\Psi^{0}_{3}\partial_{u}^{-1}\bar\sigma \right)\D\lambda\D\bar\lambda\, .
    \end{split}
\end{equation} Notably, we find new expressions for the mass and angular momentum in general relativity, involving the field $h$ which is natural from a twistor space perspective. These charges have the virtue of being fully compatible with the whole tower of $Lw_{1+\infty}$ symmetries in asymptotically flat spacetime.

\section{Discussion}
\label{sec:Discussion}

In this work, we have shown that twistor space provides the natural framework to understand the $Lw_{1+\infty}$ symmetries and their associated charges. The crucial advantage of twistor space compared to space-time is that the action of the symmetries becomes local and geometric, emerging from the gauge symmetries of the twistor space action.  This allows for the application of standard covariant phase space methods. This analysis allowed us to understand better the organization of the solutions of gravity in asymptotically flat spacetimes. 

Several questions remain to be understood. First, in this work, we essentially focused on the self-dual sector of gravity for which the twistor actions take a very simple form. Although at $\scri$, the asymptotic solution space of self-dual gravity is identical to that of gravity, it will start to differ in the subleading orders of the expansion. It would be interesting to understand where this difference exactly occurs and see to what extent the $Lw_{1+\infty}$ symmetries survive beyond the self-dual sector. We remark that a twistor action for full gravity was found by Sharma in \cite{Sharma:2021pkl}, but being based on reducing the twistor action for conformal gravity of \cite{Mason:2005zm}, it  does not sit so well with the self-dual twistor action of \cite{Mason:2007ct} used  here, see  also \cite{Mason:2009afn, Herfray:2016qvg} for related discussion.

Second, the gauge fixing conditions adopted here are quite strong and allow for a direct relation between twistor space and $\mathscr I$ via the Bramson-Tod formula. The drawback is that the expression of the residual gauge parameters becomes cumbersome involving field dependence, which requires the use of very refined tools, such as a modified Lie bracket. Another route requiring less gauge fixing and using the Penrose transform to relate twistor space data to the bulk space-time should also exist. We plan to return to this alternative approach in future work.

Third, we have not been fully explicit about the recursion operator and bihamiltonian structures implicitly used in this work, see chapter 6 of \cite{Mason:1991rf} for an introduction and more complete references.  These are common features of completely integrable partial differential equations, such as the Korteweg de Vries or nonlinear Schrödinger equations, in which there is an infinite hierarchy of conserved quantities leading to the complete integrability in the Liouville sense.  Each conserved quantity generates its own flow or higher symmetry, leading to a hierarchy of PDEs compatible with each other and the original PDE, and these can be understood to be related to each other by a recursion operator.  This recursion operator is particularly simple on twistor space where it is given in \cite{Mason:1991rf,Dunajski:2000iq} by multiplication of a \v Cech representative by the spectral parameter or ratio of twistor coordinates  $\lambda_1/\lambda_0$. Its action on perturbations of the fields is then simply obtainable from the Penrose transform integral formula using the Lax pair.  In this work, it is easily seen from our extension of the Bramson-Tod formula that in adapting the construction to $\scri$, we are instead realizing the action of the recursion operator by multiplying \v Cech, or Dolbeault representatives by a different twistor coordinate $q$. 
The action of this recursion is then easily transferred to fields on $\scri$ using the Lax pair adapted to $\scri$ as we have seen from our argument for the various recursion relations such as \eqref{recursions}.  
In the first instance, then, these are the expected symmetries of the self-dual vacuum equations as follows from their complete Liouville integrability as by now described by many authors, see  \cite{Boyer:1985aj,Mason:1991rf,Dunajski:2000iq} and references therein.  

Such symmetries are not expected to extend to the full Einstein vacuum equations in the bulk which are not believed to be integrable. There is therefore no direct evidence that these should be symmetries of the S-matrix for gravity save that they act naturally at the boundary. Indeed, the integrability of the self-dual sector implies much more than symmetry of the S-matrix,  it implies the vanishing of all scattering amplitudes except for the 3-point $\overline{\text{MHV}}$ amplitude and the 1-loop all positive external helicity amplitude \cite{Mason:2009afn}. So, the lack of integrability does not preclude some natural action on the amplitude, although it is expected to be more complicated than the Ward identities arising from the extended BMS symmetries at low $s$.

Finally, in the current paper, we focus on the celestial symmetries of gravity in asymptotically flat spacetimes. Of course, there are several possibilities for extensions. One obvious direction would be to extend this analysis to Yang-Mills theory and the $S$-algebra. Another would be to understand the effect of a cosmological constant on the whole construction and see if we can recover the recently-found deformations of $Lw_{1+\infty}$ in the presence of a cosmological constant \cite{Taylor:2023ajd,Bittleston:2024rqe}. We plan to address these questions in an upcoming paper.

\paragraph{Acknowledgements} We would like to thank Nicolas Cresto, Laurent Freidel and Marc Geiller for insightful discussions on their work. We would also like to thank Tim Adamo, Giuseppe Bogna, and Simon Heuveline for helpful discussions.  R.R. is also grateful to Glenn Barnich for related discussions during his Master's thesis project in 2016. A.K. is supported by the STFC. R.R. is supported by the Titchmarsh Research Fellowship at the Mathematical Institute and by the Walker Early Career Fellowship in Mathematical Physics at Balliol College. A.Y.S. is supported by the STFC grant ST/X000761/1. This work is supported by the Simons Collaboration on Celestial Holography.

\appendix

\section{\texorpdfstring{Recursion relation for $f_s$}{Recursion relation for fs}}
\label{sec:Recursion relation for $f_s$}

Here we derive general formulae for the $ \xi_{s}$ which are defined in terms of a field-independent part $\tilde\xi_{s}$ and a field-dependent part $f_{s}$
\begin{equation}
     \xi_{s} = \tilde \xi_{s} + f_{s}[\tau_{s},\bar\sigma]\, .
\end{equation}
First, consider the deformed complex structure acting on the field-independent part of $\xi_{s}$
\begin{align}
    [\bar \nabla\tilde \xi_s]_{\D\bar\lambda} = \sigmab \frac{u^s}{s!}\eb^{s} \tau_s - \frac{u^{s+1}}{(s+1)!}\eb^{s+2}\tau_s +(s+1)q^s \tau_s + \sigmab \sum_{n=1}^{s-1} \frac{n+1}{(s-n)!} u^{s-n}q^n \eb^{s-n}\tau_s\, .
\end{align}
We wish to add $f_s$ to this such that it cancels all the $q$ dependence. To this end, we make the Ansatz
\begin{align}
    &f_s = \sum_{n=0}^s \alpha_n\left(\lambda, \bar{\lambda}, \sigmab, u\right) q^n \\
    &\qquad \qquad \nonumber \implies [\bar \nabla f_s]_{\D\bar\lambda} = -\eb \alpha_0 +\alpha_1 \sigmab +q^{s+1} \partial_u \alpha_s + q^s \left(\partial_u \alpha_{s-1} - \eb \alpha_s \right)
    \\ &\qquad \qquad \qquad\qquad \qquad \qquad \qquad+ \sum_{n=1}^{s-1} q^n \left(\partial_u \alpha_{n-1} - \eb \alpha_n + \alpha_{n+1} \left(n+1\right) \sigmab \right)\, .
\end{align}
We will leave the dependence of the $\alpha_n$ implicit from here on. The absence of $q^{s+1}$ and $q^s$ implies
\begin{align}
    \partial_u \alpha_s = 0, \qquad \partial_u \alpha_{s-1} - \eb \alpha_s + \sigmab \left(s+1\right) \tau_s = 0 \implies \alpha_{s-1} = -\left(s+1\right) \tau_s \partial_u^{-1} \sigmab\, .
\end{align}
We have set $\alpha_s = 0$ in writing the last equation. The rest of the $\alpha_n$ must satisfy the following recursion relation $(n=s-1, \dots 1)$
\begin{align}
\label{eq:recursion}
    \partial_u \alpha_{n-1} - \eb \alpha_n + \left(n+1\right) \sigmab\alpha_{n+1}  + \frac{\left(n+1\right)}{\left(s-n\right)!}u^{s-n} \sigmab\eb^{s-n}\tau_s  = 0\, .
\end{align}
We can immediately solve for 
\begin{align}
    \alpha_{s-2} &= \partial_u^{-1} \eb\alpha_{s-1} - \frac{s}{(s-n)!}\partial_u^{-1}\left(u \sigmab\eb \tau_s \right) \\
    \nonumber &= -(s+1)\left(\partial_u^{-1}\eb\right)\left( \tau_s \partial_u^{-1} \sigmab\right)- \frac{s}{(s-n)!}\partial_u^{-1}\left(u \sigmab\eb \tau_s \right).
\end{align}
These expressions are exact and these two are only \emph{linear} in $\sigmab$. In general, we have the following expansion
\begin{align}
    \alpha_n = \sum_{p=1}^{\lfloor \frac{s-n+1}{2} \rfloor} \beta_{n,p}, \qquad \text{where } \beta_{n,p} \text{ is of degree p in } \sigmab  ,
\end{align}
and $\lfloor n \rfloor$ is the floor function. The upper limit on the sum is perhaps not obvious at this stage. However, we will derive this after solving the recursion. We can now split \eqref{eq:recursion} into different sets of recursion relations, each of which is of a fixed degree in $\sigmab$. We display the $p=1$ and $p>1$ cases separately for convenience. 
\begin{align}
        \label{eq:p1recursion}\partial_u \beta_{n-1, 1} - \eb \beta_{n,1} + \frac{\left(n+1\right)}{\left(s-n\right)!}u^{s-n} \sigmab\eb^{s-n}\tau_s  = 0, \qquad &p=1, \\
     \label{eq:higherprecursion} \partial_u \beta_{n-1, p} - \eb \beta_{n,p} + \sigmab (n+1) \beta_{n+1,p-1} = 0, \qquad &p > 1\, .
\end{align}
The solution to the $p=1$ recursion can be found by inspection to be  
\begin{align}
    \beta_{n,1} &=-\sum_{r=0}^{s-n-1} \frac{(s-r+1)}{r!} \left(\partial_u^{-1}\eb\right)^{s-n-1-r} \left(\eb^{r}\tau_s \partial_u^{-1}\left(u^{r}\sigmab\right)\right),\\
    &\nonumber =-\sum_{r=0}^{s-n-1} \frac{(s-r+1)}{r!}\mathcal{D}_{s-n,r-2}\left(\eb^{r}\tau_s u^{r}\right).
\end{align}
In the second equality above, for increased readability, we have introduced the operator $\mathcal{D}_{n,r}$ which has the following action: 
\begin{align}
\label{eq:curlyd}
    \mathcal{D}_{n,r} f = \left[\left(\partial_u^{-1}\right)^{n-2-r} \eb^{n-3-r}\right] \left(\sigmab f\right).
\end{align}
Having obtained the $\beta_{n,1}$, we can use these as seeds in \eqref{eq:higherprecursion} and use it to derive the following recursion relation in $p$:
\begin{align}
\label{eq:precursion}
    \beta_{n,p} = -\sum_{r=-2}^{s-n-3} (s-r-1) \mathcal{D}_{s-n,r}\beta_{s-r-1,p-1} \quad , \quad \beta_{n+1,0} := \frac{u^{s-n}}{(s-n)!}\bar\eth^{s-n}\tau_{s}\, .
\end{align}
The solution to this recursion can be compactly expressed using the operator \eqref{eq:curlyd} as
\begin{align}
    \beta_{n,p} &= (-1)^{p-1} \sum_{r_1=0}^{s-n-3}
    \sum_{r_2=0}^{r_1-2} \dots \sum_{r_{p-1} = 0}^{r_{p-2}-2} \prod_{i=1}^{p-1} (s-r_i-1)\mathcal{D}_{s-n,r_1}\prod_{k=1}^{p-2}\mathcal{D}_{r_k+1, r_{k+1}} \left[\beta_{s-r_{p-1}-1,1}\right]\nonumber \\
    &\nonumber =(-1)^{p} \sum_{r_1=0}^{s-n-3}
    \sum_{r_2=0}^{r_1-2} \dots \sum_{r_p = 0}^{s-n-1} \frac{1}{r_p!}\prod_{i=1}^{p} (s-r_i-1)\mathcal{D}_{s-n,r_1}\left[\prod_{k=1}^{p-2}\mathcal{D}_{r_k+1, r_{k+1}}\right]\mathcal{D}_{s-n,r_p-2} \left(\eb^{r_p} \tau_s u^{r_p}\right)\, .
\end{align}
It is now manifest that $\beta_{n,p}$ is of degree $p$ in $\sigmab$ since it has $p$ insertions of the operator $\mathcal{D}_{n,r}$. We can also infer the following limit on the value of $p$ for fixed $n, s$ from this formula. Note that the ranges on the sums imply $0 \leq r_k \leq s-n-(2k+1)$. In particular, we have $0 \leq r_{p-1} \leq s-n-2p+1$. For this to be non-empty, we must have $p \leq \lfloor\frac{1}{2}\left(s-n+1\right)\rfloor$. Thus, for a fixed spin, the gauge parameter must be
\begin{equation}
    \label{eq:finalparameter}
    \xi_s = \tilde\xi_s + \sum_{n=0}^s \sum_{p=1}^{\lfloor \frac{s-n+1}{2} \rfloor}\beta_{n,p} q^n \, .
\end{equation}
It can be checked that this formula is consistent with \eqref{field dependent f}. With these explicit forms of the gauge parameters in hand, one can check that they satisfy the algebra \eqref{Lw1infinity}. While we have not shown this analytically, we have checked explicitly that the algebra is satisfied (including all the corrections that are non-linear in $\sigmab$) for all pairs of spins up to
\begin{equation}
    \left\lbrace \tilde{\xi}_4, \tilde{\xi}_2\right\rbrace_{*}=\tilde{\xi}_5.
\end{equation}

\section{Symmetry algebra computation}\label{sec:Symmetry algebra computation}

Here we show that the field-independent part $\tilde \xi_s$ of the generators $\xi_s$ closes under the bracket \eqref{poisson bracket} when restricted to the wedge sub-algebra. We begin with
\begin{equation}
\begin{split}
    \tilde\xi_{s+s'-1} &= \sum_{n=0}^{s+s'}\frac{u^{s+s'-n}q^{n}}{(s+s'-n)!}\bar\eth^{s+s'-n}((s'+1)\tau_{s'}\bar\eth\tau_{s} - (s+1)\tau_{s}\bar\eth\tau_{s'})\\
    & = \sum_{n=0}^{s+s'}(s'+1)\frac{u^{n}q^{s+s'-n}}{n!}\bar\eth^{n}(\tau_{s'}\bar\eth \tau_{s})- (s\leftrightarrow s')\\
    &= \sum_{n=0}^{s+s'}\sum_{m=0}^{n}(s'+1)\frac{u^{n}q^{s+s'-n}}{m!(n-m)!}\bar\eth^{m+1}\tau_{s}\bar\eth^{n-m}\tau_{s'}\, ,
\end{split}
\end{equation}
now use the double sum identity 
\begin{equation}
    \sum_{k=l}^{m}\sum_{j=l}^{k}a_{k,j} = \sum_{j=l}^{m}\sum_{k=j}^{m}a_{k,j}\, .
\end{equation}
We then obtain 
\begin{equation}
\begin{split}
    \tilde\xi_{s+s'-1}&= \sum_{m=0}^{s+s'}\sum_{n=m}^{s+s'}(s'+1)\frac{u^{n}q^{s+s'-n}}{m!(n-m)!}\bar\eth^{m+1}\tau_{s}\bar\eth^{n-m}\tau_{s'}- (s\leftrightarrow s')\\
    &= \sum_{m=0}^{s}\sum_{n=0}^{s+s'-m}(s'+1)\frac{u^{n+m}q^{s+s'-n-m}}{m!n!}\bar\eth^{m+1}\tau_{s}\bar\eth^{n}\tau_{s'} - (s\leftrightarrow s')\, ,
\end{split}
\end{equation}
where we have used the wedge condition $\bar\eth^{s+2}\tau_{s} = 0$ and re-indexed. We now impose the wedge condition on the $\tau_{s'}$ and find
\begin{equation}
    \begin{split}
        \tilde\xi_{s+s'-1}\vert_{\wedge}&= \sum_{m=0}^{s}\sum_{n=0}^{s'}\frac{(s'+1)u^{n+m}q^{s+s'-n-m}}{m!n!}\bar\eth^{m+1}\tau_{s}\bar\eth^{n}\tau_{s'} \\
        &\quad\quad\quad + \sum_{m=1}^{s}\frac{u^{s'+m}q^{s-m}}{(m-1)!s'!}\bar\eth^{m}\tau_{s}\bar\eth^{s'+1}\tau_{s'} - (s\leftrightarrow s')\, .
    \end{split}
\end{equation}
We now turn to the bracket between two $\tilde\xi_s$
\begin{equation}
\begin{split}
    \{\tilde\xi_{s},\tilde\xi_{s'}\} &= \partial_{u}\tilde\xi_{s}\partial_{q}\tilde\xi_{s'} -\partial_{u}\tilde\xi_{s'}\partial_{q}\tilde\xi_{s}\\
    & = \sum_{n=0}^{s}\sum_{m=0}^{s'}\frac{(m+1)u^{s+s'-n-m}q^{n+m}}{(s-n)!(s'-m)!}\bar\eth^{s+1-n}\tau_{s}\bar\eth^{s'-m}\tau_{s'}  - (s\leftrightarrow s')\\
    & = \sum_{n=0}^{s}\sum_{m=0}^{s'}(s'+1-m)\frac{u^{n+m}q^{s+s'-n-m}}{n!m!}\bar\eth^{n+1}\tau_{s}\bar\eth^{m}\tau_{s'}  - (s\leftrightarrow s')\\
    & = \sum_{m=0}^{s}\sum_{n=0}^{s'}\frac{(s'+1)u^{n+m}q^{s+s'-n-m}}{m!n!}\bar\eth^{m+1}\tau_{s}\bar\eth^{n}\tau_{s'} - (s\leftrightarrow s')\\
    & \quad\quad \quad-\sum_{n=1}^{s+1}\sum_{m=1}^{s'}\frac{u^{n+m-1}q^{s+s'+1-n-m}}{(n-1)!(m-1)!}\bar\eth^{n}\tau_{s}\bar\eth^{m}\tau_{s'} + (s\leftrightarrow s') \\
    & = \sum_{m=0}^{s}\sum_{n=0}^{s'}\frac{(s'+1)u^{n+m}q^{s+s'-n-m}}{m!n!}\bar\eth^{m+1}\tau_{s}\bar\eth^{n}\tau_{s'} - (s\leftrightarrow s')\\ 
    &\quad\quad\quad+ \sum_{m=1}^{s}\frac{u^{s'+m}q^{s-m}}{(m-1)!s'!}\bar\eth^{m}\tau_{s}\bar\eth^{s'+1}\tau_{s'} - (s\leftrightarrow s')\, ,
\end{split} 
\end{equation}
Therefore we have that in the wedge sub-algebra
\begin{equation}
    \{\tilde\xi_{s},\tilde\xi_{s'}\} =  \tilde\xi_{s+s'-1}((s'+1)\tau_{s'}\bar\eth\tau_{s} - (s+1)\tau_{s}\bar\eth\tau_{s'})\vert_{\wedge}\, .
\end{equation}
Outside the wedge, the RHS is given by
\begin{equation}
\begin{split}
    \tilde\xi_{s+s'-1} - \tilde\xi_{s+s'-1}\vert_{\wedge} &= \sum_{m=0}^{s-2}\sum_{n=s'+2}^{s'+s-m}(s'+1)\frac{u^{n+m}q^{s+s'-n-m}}{m!n!}\bar\eth^{m+1}\tau_{s}\bar\eth^{n}\tau_{s'}\\
    & + \sum_{m=s+1}^{s+s'}\sum_{n=0}^{s+s'-m}(s'+1) \frac{u^{n+m}q^{s+s'-n-m}}{m!n!}\bar\eth^{m+1}\tau_{s}\bar\eth^{n}\tau_{s'} - (s\leftrightarrow s')\\
    & = \sum_{m=0}^{s'-1}\sum_{n=m}^{s'-1}(s'+1)\frac{u^{s+1+n}q^{s'-1-n}}{(n-m)!(s+1+m)!}\bar\eth^{s+2+m}\tau_{s}\bar\eth^{n-m}\tau_{s'}\\
    &+\sum_{m=0}^{s-2}\sum_{n=m}^{s-2}(s'+1) \frac{u^{s'+2+n}q^{s-2-n}}{m!(s'+2+n-m)!}\bar\eth^{m+1}\tau_{s}\bar\eth^{s'+2+n-m}\tau_{s'}- (s\leftrightarrow s')\\
    & = \sum_{m=0}^{s'-1}\sum_{n=m}^{s'-1}(s'+1)\frac{u^{s+1+n}q^{s'-1-n}}{(n-m)!(s+1+m)!}\bar\eth^{s+2+m}\tau_{s}\bar\eth^{n-m}\tau_{s'}\\
    & - \sum_{m=0}^{s'-2}\sum_{n=m}^{s'-2}(s+1) \frac{u^{s+2+n}q^{s'-2-n}}{m!(s+2+n-m)!}\bar\eth^{m+1}\tau_{s'}\bar\eth^{s+2+n-m}\tau_{s}- (s\leftrightarrow s')\\
    & = \sum_{m=0}^{s'-1}\sum_{n=m}^{s'-1}\frac{(s'+1)(s+2+n-m)-(s+1)m}{m!(s+2+n-m)!} u^{s+1+n}q^{s'-1-n}
    \bar\eth^{m}\tau_{s'}\bar\eth^{s+2+n-m}\tau_{s}\\ &\hspace{60pt}-(s\leftrightarrow s')\, .
\end{split}
\end{equation}
The field-independent part of the variation of the generators is given by
\begin{equation}
\begin{split}
    \delta_{s}\xi_{s'} - \delta_{s'}\xi_{s} &= \sum_{m=0}^{s'-1}\sum_{n=m}^{s'-1} q^{s'-1-n}u^{s+1+n}\frac{(s'+1-m)(m+s)!}{s!m!(s+1+n)!}\bar\eth^{n-m}(\bar\eth^{m}\tau_{s'}\bar\eth^{s+2}\tau_{s})  - (s\leftrightarrow s')\\
    & = \sum_{m=0}^{s'-1}\sum_{n=m}^{s'-1}\sum_{k=m}^{n}(s'+1-m)\frac{(m+s)!(n-m)!q^{s'-1-n}u^{s+1+n}}{s!m!(n-k)!(k-m)!(s+1+n)!}\bar\eth^{k}\tau_{s'}\bar\eth^{s+2+n-k}\tau_{s}\\&\hspace{60pt} -(s\leftrightarrow s')\\
    & = \sum_{m=0}^{s'-1}\sum_{n=m}^{s'-1}\frac{(s'+1)(s+2+n-m)-(s+1)m}{m!(s+2+n-m)!}q^{s'-1-n}u^{s+1+n}\bar\eth^{m}\tau_{s'}\bar\eth^{s+2+n-m}\tau_{s}\\&\hspace{60pt} - (s\leftrightarrow s')\, .
\end{split}
\end{equation}
A useful sum was used
\begin{equation}
\begin{split}
    \sum_{k=0}^{m}(s'+1-k)&\frac{(s+m)!(n-m)!}{m!(k-m)!} \\
    &= ((s'+1)(s+2+n-m)-m(s+1))\frac{s!(s+1+n)!(n-m)!}{m!(s+2+n-m)!}\, ,
\end{split}
\end{equation}
which relies on the use of the Chu-Vandermonde identity
\begin{equation}
    {}_{2}F_{1}(-n,b,c;1) = \frac{(c-b)_{n}}{(c)_{n}} \quad , \quad n=0,1,2,\dots\, ,
\end{equation}
where $(a)_{n} = \Gamma(a+n)/\Gamma(a)$ is the Pochhammer's symbol. Therefore, the algebra is shown to be satisfied to leading order in the fields
\begin{equation}
    \{\xi_{s},\xi_{s'}\}_{*} = \tilde{\xi}_{s+s'-1}+\mathcal{O}(\bar\sigma)\, .
\end{equation}

\section{Čech version}
\label{sec:cech version}

The homogenous formalism has the advantage that the $u$-coordinate, under Lorentz transformations, does not transform. As consequence $u$ is a weighted-scalar valued in $\mathcal{O}(1,1)$. One can recover the transformation of $u_{B}$
\begin{equation}
    u'_{B} = \frac{u}{\langle  \lambda' \hat \lambda' \rangle} = \frac{1+z \bar z}{|az+b|^2 + |cz+d|^2} u_{B}\, ,
\end{equation}
where the homogeneous coordinates on the sphere have global transformations
\begin{equation}
    \lambda'_{\alpha} = L_{\alpha}^{\,\,\,\,\beta}\lambda_{\beta} \quad , \quad L_{\alpha}^{\,\,\,\,\beta} =\begin{pmatrix}
d & c \\
b & a 
\end{pmatrix} \in \text{SL}(2,\mathbb{C})\, .
\end{equation}
Under these transformation laws, the Edth operator is not left invariant \cite{Barnich:2021dta}:
\begin{equation}
    \eth' = \eth + \frac{[\bar\lambda | \bar L TL| \hat\lambda \rangle}{\langle \hat\lambda ' \lambda ' \rangle \langle \hat \lambda\lambda \rangle}\lambda_{\alpha}\frac{\partial}{\partial \lambda_{\alpha}}\, ,
\end{equation}
where
\begin{equation}
    [\bar\lambda | \bar L TL| \hat\lambda \rangle = \bar \lambda_{\dot \beta}\bar L_{\dot \alpha}^{\,\,\,\, \dot \beta}T^{\alpha\dot \alpha} L_{ \alpha}^{\,\,\,\,  \beta}\hat \lambda_{\beta}\, .
\end{equation}
When $L \in SU(2)$ corresponds to rotations on the sphere, this term in the transformation of $\eth$ vanishes. 
Because of this fact, as written in \eqref{Bianchi identities2}, the Bianchi identities are not manifestly Lorentz invariant. However, as a curiosity, the recursion relations can be uplifted to twistor space \cite{Bramson:1977edc} as
\begin{equation}
    \frac{\p \psi_n}{\p \mu^{\dot\alpha}}= \frac{\p \psi_{n+1}}{\p\bar\lambda^{\dot\alpha}} +(3-n)\bar\lambda_{\dot\alpha} \bar\sigma\psi_{n+2}\, .\label{Bramson-form}
\end{equation}
Here $\psi_n((\lambda_\alpha,\mu^{\dot\alpha}),\bar\lambda_{\dot\alpha})$ has weight $(-n-1,n-5)$ in $(\lambda_\alpha,\mu^{\dot\alpha})$ and $\bar\lambda_{\dot\alpha}$ respectively.
This formula expresses full Lorentz invariance. Then contraction of \eqref{Bramson-form} with $\bar\lambda_{\dot\alpha}$ gives
\begin{equation}
    \frac{\p\psi_n}{\p q}= (n-4)\psi_{n+1}\, ,
\end{equation}
which can be solved in the form
\begin{equation}
    \psi_n= \sum_{m=0}^{4-n} (-1)^{m}\binom{4-n}{m} \Psi^{0}_{n+m}  q^m\, .
    \label{Expansion}
\end{equation}
The Bramson-Tod integral formula in its original form solved \eqref{Bramson-form} via a Čech cohomology integral transform
\begin{equation}
    \psi_n=\oint  \tilde q^{4-n} g( \mu^{\dot\alpha} + \tilde q \bar\lambda^{\dot\alpha}, \lambda_\alpha, \bar \lambda_{\dot\alpha}) d \tilde q\, ,
\end{equation}
where $g$ is a $\mathcal{O}(-6)$ valued holomorphic function with respect to the deformed complex structure
\begin{equation}
    \frac{\p g}{\p\bar\lambda^{\dot\alpha}} -  \bar\lambda_{\dot\alpha}\bar \sigma    \bar\lambda^{\dot\beta}\frac{\p g}{\p\mu^{\dot\beta}}  = 0 \quad, \quad \frac{\partial g}{\partial \bar\mu^{\alpha}} = 0\, .
    \label{Holomorphicity}
\end{equation}
To see that this integral formula solves the Bianchi identities we introduce for convenience $\omega^{\dot \alpha} = \mu^{\dot \alpha} + \tilde q \bar \lambda^{\dot \alpha}$ and apply the necessary derivatives 
\begin{equation}
    \begin{split}
        \frac{\partial \psi_{n}}{\partial \mu^{\dot \alpha}} &= \oint \tilde q^{4-n}\frac{\partial g}{\partial \omega^{\dot \alpha}} d\tilde q,\\
        \frac{\partial \psi_{n+1}}{\partial \bar \lambda^{\dot \alpha}} & = \oint \tilde q^{3-n}\left. \frac{\partial g}{\partial \bar \lambda^{\dot \alpha}}\right|_{\mu} d\tilde q\\
        & = \oint \tilde q^{3-n}\left(\left. \frac{\partial g}{\partial \bar \lambda^{\dot \alpha}}\right|_{\omega} + \tilde q \frac{\partial g}{\partial \omega^{\dot \alpha}}\right) d\tilde q.\\ 
    \end{split}
\end{equation}
Therefore, combining these results and using the holomorphicity of $f$ in \eqref{Holomorphicity}
\begin{equation}
\begin{split}
    \frac{\partial \psi_{n}}{\partial \mu^{\dot \alpha}} - \frac{\partial \psi_{n+1}}{\partial \bar \lambda^{\dot \alpha}} &= -\sigma^{0} \bar \lambda_{\dot \alpha}\oint \tilde q^{3-n}\bar \lambda^{\dot \beta}\frac{\partial g}{\partial \omega^{\dot \beta}} d\tilde q\\
    &= - \sigma^{0} \bar \lambda_{\dot \alpha}\oint \tilde q^{3-n} \frac{\partial g}{\partial \tilde q} d\tilde q\\
    & = (3-n)\sigma^{0}\bar \lambda_{\dot \alpha} \psi_{n+2},
\end{split}
\end{equation}
where in the last line integration parts was used. We can also understand \eqref{Expansion} by inserting the explicit expansion of $\mu^{\dot \alpha}$ into the integral formula
\begin{equation}
    \psi_{n} = \oint (\tilde q-q)^{4-n}g(\mu_{0}^{\dot \alpha}+\tilde q\bar \lambda^{\dot \alpha}, \lambda_{\alpha},\lambda_{\dot \alpha}) d\tilde q,
\end{equation}
where we introduce $\mu_{0}^{\dot \alpha} = \frac{u\hat{\bar \lambda}^{\dot \alpha}}{\langle \hat \lambda \lambda\rangle}$ and we can define the quantities living on $\scri$
\begin{equation}
    \Psi_{n}^{0} = \oint q^{4-n} g ( \mu_{0}^{\dot\alpha} +q \bar\lambda^{\dot\alpha}, \lambda_\alpha, \bar \lambda_{\dot\alpha}) dq,
\end{equation}
which solve \eqref{Bianchi identities2}.

\section{BMS Transformations}
\label{sec:BMS Transformations}

Here we demonstrate that we can recover the BMS transformation of the components of the Weyl spinor using only the integral transform and the weights naturally present in the construction. Here we take the perspective of doing infinitesimal transformations of the coordinates on twistor space and therefore under the projection to the coordinates on $\scri$. The supertranslations are 
\begin{equation}
    \delta_{\text{ST}} \mu^{\dot \alpha} = \frac{\partial T}{\partial \bar \lambda_{\dot \alpha}} \quad, \quad \delta_{\text{ST}} \lambda^{\alpha} = 0\, ,
\end{equation}
where $T$ is valued in $\mathcal{O}(1,1)$. For superrotations we take 
\begin{equation}
    \delta_{\text{SR}} \mu^{\dot \alpha } = \frac{1}{2}\frac{\partial^{2} \bar Y }{\partial \bar \lambda_{\dot\alpha}\partial \bar\lambda^{\dot \beta}} \mu^{\dot \beta} \quad ,\quad \delta_{\text{SR}} \lambda^{\alpha} = \frac{1}{2}\frac{\partial^{2}  Y }{\partial \lambda_{\alpha}\partial \lambda^{\beta}}\lambda^{\beta}\, ,
\end{equation}
where $Y,\bar Y$ are valued in $\mathcal{O}(2,0)$ and $\mathcal{O}(0,2)$ respectively. All the variations can also be extended to $\bar \lambda_{\dot \alpha}$ and $\bar \mu^{\alpha}$ by complex conjugation. Furthermore, the BMS generators $\{T,Y,\bar Y\}$ are only functions of $\lambda_{\alpha}$ and $\bar \lambda_{\dot \alpha}$.  All variations are taken on-shell and therefore $\partial_{\bar u} g_{1} = 0$. We begin by using the Bramson-Tod integral formula and inserting the infinitesimal coordinate transformations. Starting with supertranslations
\begin{equation}
    \begin{split}
        \delta_{\text{ST}} \mathcal{Q}_{s} &= \int q^{s+2}\left(\frac{\partial g_{1}}{\partial \mu^{\dot \alpha}}\delta_{\text{ST}}\mu^{\dot\alpha} + \frac{\partial g_{1}}{\partial \bar \mu^{ \alpha}}\delta_{\text{ST}}\bar\mu^{\alpha} \right)\d q\d\bar q \\
        & = \int q^{s+2}\left(\frac{\partial g_{1}}{\partial \mu^{\dot \alpha}}\frac{\partial T}{\partial \bar \lambda_{\dot \alpha}} + \partial_{\bar q}g_{1} \eth \bar T \right)\d q\d\bar q\\
        & = \int q^{s+2}\left(T \partial_{u}g_{1} - \partial_{q}g_{1}\bar\eth T \right)\d q\d\bar q\\
        & = T \partial_{u}\mathcal{Q}_{s} + (4-n)\mathcal{Q}_{s-1}\bar\eth T\, ,
    \end{split}
\end{equation}
where integration by parts and the homogeneity of $T$ has been used. Now moving on to superrotations 
\begin{equation}
    \delta_{\text{SR}}\mathcal{Q}_{s} = \int q^{s+2}\left(\frac{\partial g_{1}}{\partial \mu^{\dot \alpha}}\delta_{\text{SR}}\mu^{\dot \alpha}+ \frac{\partial g_{1}}{\partial \bar\lambda^{\dot \alpha}}\delta_{\text{SR}}\bar\lambda^{\dot \alpha} + \frac{\partial g_{1}}{\partial \bar\mu^{ \alpha}}\delta_{\text{SR}}\bar\mu^{ \alpha}+ \frac{\partial g_{1}}{\partial \lambda^{ \alpha}}\delta_{\text{SR}}\lambda^{ \alpha}\right)\d q\d\bar q\, .
\end{equation}
Taking this term by term starting with the dotted spin bundle terms
\begin{equation}
\begin{split}
    \int q^{s+2}\frac{\partial g_{1}}{\partial \mu^{\dot \alpha}}\delta_{\text{SR}}\mu^{\dot \alpha}  \d q\d\bar q 
    &= \frac{1}{2}\int q^{s+2}\epsilon^{\dot \alpha \dot \beta} \frac{\partial g_{1}}{\partial \mu^{\dot \beta}}\left(u\bar\eth\left(\frac{\partial \bar Y}{\partial \bar \lambda^{\dot \alpha}}\right)+ q \frac{\partial \bar Y}{\partial \bar \lambda^{\dot \alpha}}\right) \d q\d\bar q\\
    &=\frac{1}{2}\int q^{s+2}\left[ \partial_{u}g_{1}\left(u\bar \eth \bar Y + 2q\bar Y\right)-\partial_{q}g_{1}\left(u\bar\eth^{2}\bar Y + q\bar \eth \bar Y\right) \right] \d q\d\bar q\\
    & = \frac{u}{2}\bar \eth \bar Y \partial_{u}\mathcal{Q}_{s} + \bar Y \partial_{u}\mathcal{Q}_{s+1} + \frac{s+3}{2}\bar \eth\bar Y \mathcal{Q}_{s} + \frac{u}{2}(s+2)\mathcal{Q}_{s-1}\bar \eth^{2}\bar Y \, ,
\end{split}
\end{equation}
where we only use the homogeneity of $\bar Y$ and integration by parts. The next term gives
\begin{equation}
\begin{split}
    \int q^{s+2} \frac{\partial g_{1}}{\partial \bar\lambda^{\dot \alpha}}\delta_{\text{SR}}\bar\lambda^{\dot \alpha} \d q\d\bar q &= \frac{1}{2}\int q^{s+2} \epsilon^{\dot \alpha \dot \beta} \frac{\partial g_{1}}{\partial \bar \lambda^{\dot \beta}}\frac{\partial \bar Y}{\partial \bar \lambda^{\dot \alpha}}\d q\d\bar q\\
    & = \frac{1}{2}\int q^{s+2}\left[ 2\bar Y \bar \partial_{0}g_{1} - \bar\lambda^{\dot \alpha}\frac{\partial g_{1}}{\partial \bar \lambda^{\dot \alpha}} \bar \eth \bar Y\right]\d q\d \bar q\\
    & = \frac{1}{2}\int q^{s+2}\left[ 2\bar Y \bar \partial_{0}g_{1} + (g_{1}+\bar u\partial_{\bar u}g_{1}+\bar q\partial_{\bar q}g_{1}) \bar \eth \bar Y\right]\d q\d \bar q\\
    & = \bar Y\int q^{s+2}\left[\bar\eth - q\partial_{u} + \frac{\bar u}{\langle \hat \lambda \lambda \rangle^{2} }\partial_{\bar q}\right]g_{1}\d q\d \bar q\\
    & = \bar Y \bar\eth \mathcal{Q}_{s} - \bar Y \partial_{u}\mathcal{Q}_{s+1}\, ,
\end{split}
\end{equation}
where we have used the fact that $g_{1}$ is a section of $\mathcal{O}(-5,-1)$ and on-shell $\partial_{\bar u}g_{1} = 0$. Now looking at the un-dotted transformations
\begin{equation}
\begin{split}
    \int q^{s+2} \frac{\partial g_{1}}{\partial\bar \mu^{ \alpha}}\delta_{\text{SR}}\bar\mu^{ \alpha}  \d q\d\bar q  &= - \frac{1}{2}\int q^{s+2}\partial_{\bar q}g_{1} \left( \bar u \eth^{2}Y+\bar q \eth Y \right)\d q\d\bar q= \frac{1}{2}\eth Y \mathcal{Q}_{s}\, ,
\end{split}
\end{equation}
where we have used the homogeneity of $Y$ and integration by parts. The final term gives us
\begin{equation}
    \begin{split}
        \int q^{s+2} \frac{\partial g_{1}}{\partial \lambda^{\alpha}}\delta_{\text{SR}}\lambda^{\alpha} \d q\d\bar q &= \frac{1}{2}\int q^{s+2}\left[2Y\partial_{0}g_{1} - \lambda^{\alpha}\frac{\partial g_{1}}{\partial \lambda^{\alpha}}\eth Y\right]\d q \d \bar q\\
        & = \frac{1}{2}\int q^{s+2}\left[\eth Y(u\partial_{u}g_{1}+(2-s)f)+2 Y\left(\eth g_{1}+ \frac{u}{\langle\hat \lambda\lambda\rangle^{2}}\partial_{q}f\right)  \right]\d q\d\bar q\\
        & = \frac{u}{2}\eth Y \partial_{u}\mathcal{Q}_{s} + \frac{2-s}{2}\eth Y \mathcal{Q}_{s} +Y \eth\mathcal{Q}_{s} - (s+2)\frac{u}{\langle \hat \lambda \lambda \rangle^{2}}Y\mathcal{Q}_{s-1}\\
        & = \frac{u}{2}\eth Y \partial_{u}\mathcal{Q}_{s} + \frac{2-s}{2}\eth Y \mathcal{Q}_{s} +Y \eth\mathcal{Q}_{s} + \frac{u}{2}(s+2)\mathcal{Q}_{s-1}[\bar \eth ,  \eth]Y\, ,
    \end{split}
\end{equation}
where we have used the homogeneity of $g_{1}$ and $\partial_{\bar u} g_{1} = 0$. A useful commutator between the $\eth$ and $\bar \eth$ operators was used
\begin{equation}
    [\eth , \bar \eth] = \frac{1}{\langle \hat\lambda \lambda \rangle^{2}}\left[\lambda_{\alpha}\frac{\partial }{\partial \lambda_{\alpha}} - \bar\lambda_{\dot\alpha}\frac{\partial }{\partial \bar\lambda_{\dot\alpha}}\right] \, .
\end{equation}
Putting all of this together, along with the assumption that $\bar\eth Y = 0$, the full infinitesimal BMS transformations give
\begin{equation}
    \delta_{(T,Y,\bar Y)} \mathcal{Q}_{s} = \left(f\partial_{u} +Y\eth + \bar Y \bar \eth + \frac{3-s}{2}\eth Y + \frac{3+s}{2}\bar\eth \bar Y\right) \mathcal{Q}_{s}+(s+2)\mathcal{Q}_{s-1}\bar \eth f  \, ,
\end{equation}
where $f = T +\frac{u}{2}\left(\eth Y + \bar \eth \bar Y\right)$.

\section{\texorpdfstring{Vectors and Forms on $\mathbb{CP}^{n}$}{Vectors and Forms on CPn}}\label{sec:ProjForms}
Due to the projective nature of twistor space, it is worth discussing the tangent and cotangent space on $\mathbb{CP}^{n}$ which is defined as
\begin{equation}
    \mathbb{CP}^{n} = \{ (Z^{0}, \cdots , Z^{n})\in \mathbb{C}^{n+1} \vert  (Z^{0}, \cdots , Z^{n}) \sim  (t Z^{0}, \cdots , t Z^{n}) : t \in \mathbb{C}^{*}\},
\end{equation}
where the $Z_{A}$ are called homogeneous coordinates. We wish to use these coordinates rather than choosing a patch $U_{i}$ where $Z_{i} \neq 0$ and the affine coordinates are $z_{j} = Z_{j}/Z_{i}$ such that $i\neq j$ to keep our formula agnostic to the choice of the patch. To construct the tangent space, we consider the space of all derivations on functions 
\begin{equation}
    f: \mathbb{CP}^{n} \rightarrow \mathbb{C},
\end{equation}
which in homogenous coordinates can be viewed as
\begin{equation}
    f: \mathbb{C}^{n+1} \rightarrow \mathbb{C} \quad, \quad f(t Z) = f(Z),
\end{equation}
a function with zero homogeneity. Therefore the homogeneity  vector field $\Upsilon = Z^{A}\partial_{A}$ which generates rescaling will not appear, put another way
\begin{equation}
    T^{1,0}_{Z}\mathbb{CP}^{n} \otimes \mathcal{O}(m, n) \cong T^{1,0}_{Z}\mathbb{C}^{n+1} / \langle \Upsilon \rangle ,
\end{equation}
where $\langle \Upsilon \rangle \subset T^{1,0}_{Z}\mathbb{C}^{n+1}$ and the twisting of the line bundle $\mathcal{O}(m, n) $ appears as our vector fields may have a non zero homogeneity. Therefore to get sections of  $T^{1,0}_{Z}\mathbb{CP}^{n}$ we must multiply our vector fields by sections of $\mathcal{O}(-m, -n) $. The cotangent bundle is realized by $\omega \in \Omega^{1,0}(\mathbb{C}^{n+1})$ such that $\Upsilon \lrcorner \, \, \omega = 0$. More generally sections of $\Omega^{p,q}(\mathbb{CP}^{n})\otimes \mathcal{O}(m, n) $ are given by the subspace of $\Omega^{p,q}(\mathbb{C}^{n+1})$ such that \eqref{homogene} are satisfied. The next task is to define the exterior derivative on $\mathbb{CP}^{n}$ from the one on $\mathbb{C}^{n+1}$, as these two objects do not coincide due to the presence of the line bundle. By definition of a connection on a vector bundle, we expand the exterior derivative acting on forms $ \omega$ valued in $\mathcal{O}(m,n) $ on $\mathbb{CP}^{n}$ as
\begin{equation}
    d_{\mathbb{CP}^{n}} \omega  = \d\omega  - m \theta \wedge\omega - n\bar \theta \wedge\omega,
\end{equation}
where $d  = \d Z^{A}\frac{\partial}{\partial Z^{A}} + \d\bar Z^{\bar A}\frac{\partial}{\partial \bar Z^{\bar A}}$. By requiring that $d_{\mathbb{CP}^{n}} \omega$ is a form under the conditions \eqref{homogene}, we find that 
\begin{equation}
    \Upsilon \lrcorner \, \, \theta = 1 \quad ,\quad \bar\Upsilon \lrcorner \, \, \bar\theta = 1,
\end{equation}
and all other contractions are zero. This means that the connection $1$-forms must lie in the image of the homogeneity vector fields, and therefore can be built in the process of building the cotangent space. As an example, let us take the canonical bundle of $\mathbb{CP}^{n}$ which is defined as $K_{\mathbb{CP}^{n}} = \bigwedge^{n} T^{* 1,0} \mathbb{CP}^{n}$. Taking an n-form on $\mathbb{C}^{n+1}$
\begin{equation}
    \omega = \sum_{A = 0}^{n} f_{A}(Z) \d Z^{0} \wedge \cdots \wedge \hat{\d Z^{A}} \wedge \cdots \wedge \d Z^{n}, 
\end{equation}
where $\hat{dZ^{A}}$ corresponds to the omission. Now taking the contraction with the homogeneity vector field
\begin{equation}
    \Upsilon \lrcorner \, \, \omega  = \sum_{A,B = 0}^{n} (-1)^{B} f_{A} (Z) Z_{B} \d Z^{0} \wedge \cdots \wedge \hat{\d Z^{A}} \wedge \cdots \wedge \hat{\d Z^{B}} \wedge \cdots \wedge \d Z^{n},
\end{equation}
which vanishes if $f_{A}(Z) = (-1)^{A} Z_{A} g(Z)$ by antisymmetry. Therefore we can compactly write the forms in $K_{\mathbb{CP}^{n}}$ as
\begin{equation}
    \omega = g(Z) \epsilon_{A_{0}\cdots A_{n}} Z^{A_{0}} \d Z^{A_{1}}\wedge \cdots \wedge \d Z^{A_{n}},
\end{equation}
with $g(Z)$ being a section of $\mathcal{O}(-n-1)$ and therefore $K_{\mathbb{CP}^{n}} \cong \mathcal{O}(-n-1) $. Another example is provided by the simplest case of $\mathbb{CP}^{1}$, which sees much use in twistor theory. The unique sections of $T^{ 1,0} \mathbb{CP}^{1}\otimes \mathcal{O}(-2)$ and $T^{* 1,0} \mathbb{CP}^{1} \otimes \mathcal{O}(2)$ are given by
\begin{equation}
    [\eth]= \left[ \frac{a^{\alpha}}{\langle a \lambda \rangle } \frac{\partial}{\partial \lambda^{\alpha}}\right] \quad \quad e^{0} = \D\lambda = \langle \lambda \d\lambda\rangle ,
\end{equation}
where $[a] \in \mathbb{CP}^{1}$ and $[\lambda ]\neq [a]$ is a reference spinor. The definition of $[\eth]$ is independent of the spinor $a$ due to the equivalence class and can be defined globally. The exterior derivative acting on forms valued in $\mathcal{O}(m,n)$ is given by
\begin{equation}
    d_{\mathbb{CP}^{1}} = d - m\frac{\langle a \d\lambda \rangle}{\langle a \lambda \rangle} \wedge - n\frac{[\bar a \d \bar\lambda ]}{[\bar a \bar\lambda]}\wedge.
\end{equation}
On twistor space, we choose a decomposition that is adapted to $\scri$, in the notation introduced in \eqref{Basis} 
\begin{equation}
    \Omega^{1}(\mathbb{PT}) = \text{Span}\left\{\D\lambda , \partial \mathfrak{q}, \partial u_{B}, \bar e^{0}, \bar\partial \bar{\mathfrak{q}}, \bar\partial \bar u_{B}\right\}  \quad, \quad \theta = \frac{\langle\hat\lambda  \d\lambda\rangle}{\langle\hat\lambda \lambda \rangle} \quad, \quad \bar \theta = \frac{\langle \d\hat\lambda \lambda\rangle}{\langle\hat\lambda\lambda\rangle}\, .
\end{equation}
Note that the connections $\theta, \bar\theta$ pull back to the same expression on $\scri$ and is a particularly natural choice. The holomorphic top form on twistor space can be decomposed into the coordinate of the fibration over $\scri$
\begin{equation}
\begin{split}
    \D^{3}Z &=\frac{1}{4!} \epsilon_{ABCD}Z^{A}\d Z^{B}\wedge\d Z^{C}\wedge \d Z^{D}\\
    & = \frac{1}{2}\D\lambda\wedge \d\mu^{\dot \alpha}\wedge \d \mu_{\dot \alpha} + \frac{1}{2}[\mu\d \mu]\wedge \d\lambda^{\alpha}\wedge \d\lambda_{\alpha}\\
    & = \d_{\mathbb{CP}^{1}}u \wedge \d_{\mathbb{CP}^{1}} q \wedge \D\lambda + q \d_{\mathbb{CP}^{1}}q\wedge \D\bar\lambda \wedge\D\lambda\\
    & = \langle\lambda \hat\lambda\rangle\d u_{B}\wedge \d_{\mathbb{CP}^{1}}q\wedge \D\lambda + q \d_{\mathbb{CP}^{1}}q\wedge \D\bar\lambda \wedge\D\lambda\, .
\end{split} 
\end{equation} 
Once we restrict to a constant $u_{B}$ the first term vanishes. Note that since the decomposition of $\mu^{\dot\alpha}$ does not preserve the complex structure, the presence of a $\D\bar\lambda$ form is to be expected. We also note that $\d_{\mathbb{CP}^{1}}\bar q = \langle \lambda\hat\lambda\rangle\bar\partial \bar{\mathfrak{q}}$ and therefore the restriction of the form in the Bramson-Tod formula can be equally written as
\begin{equation}
    g\vert_{X_{(u,\lambda,\bar\lambda)}} = g_{1} \d_{\mathbb{CP}^{1}} \bar q= g_{\bar{\mathfrak{q}}}\bar\partial \bar{\mathfrak{q}}\, ,
\end{equation}
in the notation of Section \ref{sec:Asymptotic twistor space and Tod's integral formula}. We drop the $\mathbb{CP}^{1}$ subscript for notational convenience in the rest of the text.

\addcontentsline{toc}{section}{References}
\bibliographystyle{style}
\bibliography{Biblio}

\end{document}